\documentclass[twocolumn,trackchanges]{aastex631}
\usepackage[utf8]{inputenc}   
\usepackage[T1]{fontenc}      
\usepackage{lmodern}          

\usepackage{amsmath}
\usepackage{threeparttable} 
\usepackage{tikz}
\usetikzlibrary{shapes,arrows}
\usepackage{float}
\usepackage{hyperref}
\usepackage{graphicx}
\usepackage{multirow}
\usepackage{tabularx}
\usepackage[figuresright]{rotating}
\usepackage{booktabs} 
\graphicspath{{./}{figures/}}
\usepackage{xcolor}
\usepackage{listings}

\lstdefinelanguage{json}{
  basicstyle=\ttfamily\footnotesize,
  numbers=none,
  breaklines=true,
  breakatwhitespace=false,
  columns=fullflexible,
  showstringspaces=false,
  morestring=[b]",
  morecomment=[l]{//},
  stringstyle=\color{blue!60!black},
  commentstyle=\color{gray},
  literate=
   *{,}{{,\allowbreak}}{1}
    {<}{{<\allowbreak}}{1}
    {>}{{>\allowbreak}}{1}
    {_}{{\_\allowbreak}}{1}
    {0}{{{\color{blue}0}}}{1}
    {1}{{{\color{blue}1}}}{1}
    {2}{{{\color{blue}2}}}{1}
    {3}{{{\color{blue}3}}}{1}
    {4}{{{\color{blue}4}}}{1}
    {5}{{{\color{blue}5}}}{1}
    {6}{{{\color{blue}6}}}{1}
    {7}{{{\color{blue}7}}}{1}
    {8}{{{\color{blue}8}}}{1}
    {9}{{{\color{blue}9}}}{1}
}

\shorttitle{A review of AGB star}
\shortauthors{Lu et al.}

\begin{document}
\footnotetext{$^{\dagger}$ These authors contributed equally to this work.}

\title{Spectra as Language: Large Language Models for Scalable Stellar Parameter and Abundance Inference}

\correspondingauthor{Yin-Bi Li \& Cun-Shi Wang \& A-Li Luo}
\email{* ybli@bao.ac.cn, wangcunshi@nao.cas.cn, lal@nao.cas.cn}

\author[0000-0002-8252-8743]{Hai-Ling Lu$^{\dagger}$\thanks{These authors contributed equally to this work.}}
\affiliation{National Astronomical Observatories, Chinese Academy of Sciences, Beijing 100101, China}
\affiliation{University of Chinese Academy of Sciences, Beijing 100049, China}

\author[0009-0004-5525-2982]{Yu-Yang Li$^{\dagger}$\thanks{These authors contributed equally to this work.}}
\affiliation{National Astronomical Observatories, Chinese Academy of Sciences, Beijing 100101, China}
\affiliation{School of Astronomy and Space Science, University of Chinese Academy of Sciences, Beijing 100049, China}

\author[0000-0001-7607-2666]{Yin-Bi Li$^{*}$}
\affiliation{National Astronomical Observatories, Chinese Academy of Sciences, Beijing 100101, China}
\affiliation{University of Chinese Academy of Sciences, Beijing 100049, China}

\author[0000-0001-5941-3246]{Cun-Shi Wang$^{*}$}
\affiliation{National Astronomical Observatories, Chinese Academy of Sciences, Beijing 100101, China}
\affiliation{School of Astronomy and Space Science, University of Chinese Academy of Sciences, Beijing 100049, China}

\author[0000-0001-7865-2648]{A-Li Luo$^{*}$}
\affiliation{National Astronomical Observatories, Chinese Academy of Sciences, Beijing 100101, China}
\affiliation{School of Astronomy and Space Science, University of Chinese Academy of Sciences, Beijing 100049, China}
\affiliation{University of Chinese Academy of Sciences, Nanjing 211135, China}

\author[0009-0001-9085-8718]{Jun-Chao Liang}
\affiliation{National Astronomical Observatories, Chinese Academy of Sciences, Beijing 100101, China}
\affiliation{University of Chinese Academy of Sciences, Beijing 100049, China}

\author[0000-0002-8913-3605]{Shuo Li}
\affiliation{National Astronomical Observatories, Chinese Academy of Sciences, Beijing 100101, China}
\affiliation{University of Chinese Academy of Sciences, Beijing 100049, China}

\begin{abstract}

Stellar spectra encode key information on the physical properties and chemical compositions of stars. Accurate determination of stellar parameters constitutes a fundamental cornerstone of astrophysical research and is crucial for addressing major scientific questions such as galaxy evolution and stellar evolution. With the rapid advancement of large-scale stellar spectroscopic surveys, an unprecedented volume of spectral data has been accumulated. Traditional approaches based on handcrafted feature extraction or physical model fitting face challenges when dealing with high-dimensional, massive spectral datasets, including limited generalization capability, low computational efficiency, and difficulty in achieving the simultaneous and precise estimation of multiple parameters. In recent years, large language models have demonstrated strong generalization and feature-learning capabilities in sequence-related tasks such as natural language processing, DNA/RNA sequence analysis, and protein and chemical formula parsing. Given that stellar spectra are intrinsically continuous sequential signals, the transferability of large language models to stellar spectroscopy has become a valuable research direction.
In this work, we propose a two-stage trained large language model framework for stellar parameter determination. The framework enables robust and accurate inference of fundamental stellar atmospheric parameters, including effective temperature, surface gravity, and overall metallicity, and, building upon this foundation, achieves accurate joint estimation of the abundances of nearly twenty chemical elements. Furthermore, scaling-law analyses based on spectroscopic data augmentation demonstrate a stable and systematic improvement in model performance with increasing data scale, providing a scalable and efficient technical framework for precise parameter determination in forthcoming large-scale stellar spectroscopic surveys.

\end{abstract}


\section{Introduction} \label{sec:intro}

Stellar spectra are often regarded as “cosmic codes” for decoding the physical nature of stars, encoding key information such as the effective temperature (\textit{T$_\mathrm{eff}$}), surface gravity (\textit{log\,$g$}), metallicity ([Fe/H]), and the abundances of various chemical elements. They constitute a fundamental tool for revealing the laws of stellar formation and evolution, tracing the history of galactic chemical enrichment, and probing the origin of cosmic structures. With astronomy entering the era of large-scale surveys, major projects such as LAMOST and APOGEE have accumulated massive volumes of stellar spectra—LAMOST alone has released more than ten million low-resolution spectra \citep{2012RAA....12.1197C}, while APOGEE has provided several hundred thousand high-precision infrared spectra \citep{2022ApJS..259...35A, 2025arXiv251104365S}. The rapid growth of spectroscopic data offers unprecedented opportunities for comprehensively reconstructing the evolutionary picture of the Milky Way and the Universe, while simultaneously posing significant challenges to efficient and accurate stellar parameter determination.

Current mainstream approaches to stellar parameter estimation can be broadly categorized into two classes. The first class consists of physics-based fitting models, such as The Payne and ASPCAP \citep{2008AJ....136.2022L, 2011AJ....141...89S, 2015ApJ...808...16N,2016AJ....151..144G, 2017MNRAS.464.3657X, 2019ApJ...879...69T,2019ApJS..245...34X,2020AJ....160...83S}. These methods rely on detailed assumptions of stellar atmosphere physics and infer parameters by fitting differences between theoretical and observed spectra. However, they have notable limitations: on the one hand, complex physical assumptions are difficult to adapt to all stellar types (e.g., chemically peculiar stars or stars at late evolutionary stages); on the other hand, the fitting procedures are computationally expensive, making them inefficient for large-scale survey data. The second class includes data-driven deep learning approaches, such as convolutional neural networks (CNNs) and recurrent neural networks (RNNs) \citep{2020MNRAS.498.3817B, 2020PASP..132b4504Z,2021ApJ...906..130O, 2022MNRAS.517.4875L,2025ApJS..281...58L,2025RASTI...4af048G}. These methods automatically extract spectral features and improve prediction efficiency. Nevertheless, existing deep learning models generally suffer from a lack of task generality: different parameters (e.g., basic atmospheric parameters versus heavy-element abundances) or different spectral regimes (e.g., low- versus high-resolution spectra) often require dedicated network architectures and extensive retraining, resulting in poor transferability and limited generalization. Moreover, the 
“black-box” nature of these models reduces interpretability, constraining their broader adoption in frontier astrophysical research.

In recent years, large language models have demonstrated remarkable generalization and feature-learning capabilities across various sequence-related tasks, including natural language processing, DNA/RNA sequence analysis, and protein and chemical formula parsing \citep{2023arXiv230813565X, 2024arXiv240114656Z}. Their core strength lies in learning universal representations from massive and heterogeneous sequence data without redesigning model architectures for specific tasks. Against this backdrop, large language models (LLMs) have achieved breakthroughs in natural language processing and multimodal tasks, with their application scope gradually extending beyond text and vision to more general numerical sequence modeling problems. For numerical sequence tasks, \cite{2024arXiv240201801Z} systematically reviews recent efforts to bridge the modality gap between text-trained LLMs and continuous time-series data, summarizing representative strategies such as prompting-based textualization, numerical quantization, representation alignment, vision-based bridging, and tool integration for time-series analysis.

From a complementary engineering and application-oriented perspective, \cite{2024arXiv240203182J} further reviews the general workflow and key challenges of LLM-based time-series analysis. Compared with training generic time-series models from scratch, leveraging the representational power of pretrained LLMs through suitable tokenization, prompt design, and fine-tuning strategies can offer greater flexibility for large-scale, diverse, and non-stationary time-series tasks. This review also highlights the prevalence of distribution shift and concept drift in time-series problems, which impose additional challenges on continual adaptation and cross-dataset generalization, underscoring the importance of constructing unified and transferable sequence modeling frameworks. Together, these two surveys indicate that LLMs are evolving from 
“text models” into general-purpose modeling paradigms for broad classes of sequential data.

Notably, stellar spectra are also continuous numerical sequences ordered along the wavelength axis, where spectral lines and continua across different bands encode long-range physical and chemical correlations. As such, they naturally fit within the aforementioned “numerical sequence--LLM” methodological framework. In astronomy, LLM- and transformer-based studies have already achieved initial success in astronomical sequence tasks \citep{2024arXiv240410757L, 2024arXiv240410019W, 2025arXiv251108970S,2025MLS&T...6d5005Z, 2025arXiv251017960P, 2025NatAs...9.1869S}. For example, \cite{2025arXiv250810075R} demonstrated that parameter-efficient fine-tuning (LoRA) can directly adapt pretrained LLMs to spectroscopic data, enabling quantitative prediction of galaxy spectral redshifts while achieving competitive spectroscopic performance. In addition, \cite{2026ApJ..1000...14F} showed that ABC-SN outperforms the CNN-based DASH model for nearly all supernova subtypes, indicating that attention mechanisms can effectively capture discriminative spectral features and further supporting the treatment of spectra as structured sequences suitable for attention-based modeling. More recently, \cite{2026ApJ...998..189Z} proposed SpecCLIP, a unified foundation-model framework that pretrains on LAMOST and Gaia XP spectra and uses contrastive alignment along with spectrum-aware decoders to enable cross-spectrum translation and simultaneous inference of multiple stellar parameters and elemental abundances. This work represents a concrete implementation of a unified modeling approach for quantitative stellar spectroscopy.

Guided by the central idea of “spectra as language”, this work introduces the general sequence-learning capabilities of LLMs into stellar spectroscopic analysis to construct a unified modeling framework tailored for large-scale survey data. The continuous flux distributions of stellar spectra encode physical and chemical information through line structures and continuum shapes, which are analogous, at the sequence modeling level, to how discrete symbols in natural language convey semantics via contextual relationships. Specifically, we adopt a two-stage fine-tuning strategy. In the first stage, the model is trained on LAMOST low-resolution spectra with atmospheric parameter labels derived from the LAMOST pipeline, in order to build foundational capabilities and achieve accurate inference of core stellar atmospheric parameters. In the second stage, the model is trained on LAMOST low-resolution spectra that are co-origin with APOGEE observations, using chemical abundance labels transferred from APOGEE high-resolution spectroscopy, thereby extending the model capacity and enabling the simultaneous inversion of multiple elemental abundances.

To systematically present this study, the remainder of the paper is organized as follows. The Data section describes the sources of labeled data, including stellar samples from LAMOST DR11 and APOGEE DR16, along with detailed data selection and preprocessing procedures. The Methodology section then covers three core components—sample construction, model structure design, and training methods—providing a comprehensive description of the two-stage fine-tuning framework. Subsequently, the Results section presents model performance evaluations in a stage-wise manner. Finally, the Discussion section focuses on an in-depth analysis of model performance and the methodological implications of this work.

\section{ Data} \label{sec:data}

\subsection{Stellar Samples with Labels from LAMOST DR11}

The Large Sky Area Multi-Object Fiber Spectroscopic Telescope (LAMOST), also known as the Guo Shoujing Telescope, is a large optical observational facility independently designed and constructed by China, and is operated and maintained by the National Astronomical Observatories of the Chinese Academy of Sciences. LAMOST is a reflecting Schmidt telescope equipped with 4,000 fibers on its focal plane, enabling the simultaneous observation of 4,000 targets within a field of view of 20 square degrees, thereby significantly increasing the efficiency of spectroscopic data acquisition \citep{2012RAA....12.1197C, 2015RAA....15.1095L}.

As of June 2023, LAMOST has released its 11th data release (DR11). DR11 v1.1 contains a total of 11,944,049 low-resolution spectra, including 11,586,067 stellar spectra. Among these stellar spectra with available atmospheric parameter labels, there are 432,474 A-type stars, 2,252,262 F-type stars, 3,712,147 G-type stars, 1,403,434 K-type stars, and 899,204 M-type stars. The spectra cover a wavelength range from 3,700~\AA\ to 9,000~\AA\ with a resolving power of approximately $R \sim 1{,}800$. 


To construct datasets of different scales, we perform balanced subsampling across stellar spectral types, with an equal number of samples drawn from each class. For example, for the dataset of $10^5$ samples, approximately $2\times10^4$ stars were selected for each of the A, F, G, K, and M spectral types. For the dataset of $1.5\times10^6$ samples, approximately $3\times10^5$ stars were selected for each spectral type.

\subsection{Stellar Samples with Labels from APOGEE DR16}

The Apache Point Observatory Galactic Evolution Experiment (APOGEE) is one of the major sub-projects of the Sloan Digital Sky Survey phases III (SDSS-III) and IV (SDSS-IV). APOGEE is a large-scale stellar spectroscopic survey conducted in the near-infrared H band. It employs two fiber-fed cryogenic spectrographs mounted on the 2.5\,m Sloan Foundation Telescope at Apache Point Observatory (APO) in New Mexico, USA, and on the 1\,m NMSU telescope at Las Campanas Observatory (LCO) in Chile. APOGEE is distinctive among large surveys in that it provides high-resolution spectra ($R \approx 22{,}500$) covering the entire Milky Way.

One of the primary scientific goals of APOGEE is to derive multi-element chemical abundances for a large stellar sample. The APOGEE Stellar Parameters and Abundances Pipeline (ASPCAP) is a dedicated processing pipeline developed for APOGEE spectra, which analyzes stellar spectra to derive atmospheric parameters and individual chemical element abundances. With successive data releases (DRs), the algorithms, models, and performance of ASPCAP have been continuously improved and refined.

In this work, we adopt the multi-element abundance parameters provided by the APOGEE DR16 pipeline as supervised learning labels. The selected parameters include abundance ratios for a variety of elements, such as C, N, O, Na, Mg, Al, Si, P, S, K, Ca, Ti, V, Cr, Mn, Co, Ni, and Cu, all expressed in the form of [X/Fe].

Since APOGEE spectra and derived parameters may suffer from potential quality issues, it is necessary to apply the bitmask flags provided by APOGEE to select reliable data. The \texttt{ASPCAPFLAG} bitmask is used to indicate possible issues in the ASPCAP fitting process. In this study, we required \texttt{ASPCAPFLAG} to be set to 0 and simultaneously enforce all relevant parameter-level flag fields to be 0. In addition, atmospheric parameter entries with the invalid placeholder value of −9999.99 were ignored, thereby removing potential quality issues and ensuring the reliability of the training and analysis datasets. The resulting sample counts for each parameter are summarized in Table~\ref{tab:Valid Parameter Count_with_APOGEE}.

\begin{table*}[ht!]
\centering
\caption{Number of spectra with valid elemental abundance labels for each stellar spectral class}
\renewcommand{\arraystretch}{1.2}

\begin{tabular}{lccccccc}
\hline\hline
\textbf{Class}
  & \textbf{C\_FE} & \textbf{CI\_FE} & \textbf{N\_FE} & \textbf{O\_FE}
  & \textbf{MG\_FE} & \textbf{AL\_FE} & \textbf{SI\_FE} \\
  & \textbf{S\_FE} & \textbf{MN\_FE} & \textbf{NI\_FE} & \textbf{CA\_FE}
  & \textbf{NA\_FE} & \textbf{P\_FE} & \textbf{K\_FE} \\
  & \textbf{TI\_FE} & \textbf{TIII\_FE} & \textbf{CR\_FE} & \textbf{CO\_FE}
  & \textbf{CU\_FE} & \textbf{V\_FE} & \textbf{CE\_FE} \\
\hline

A & 441 & 441 & 330 & 92  & 569 & 554 & 567 \\
  & 535 & 538 & 563 & 474 & 131 & 95  & 178 \\
  & 132 & 148 & 171 & 145 & 177 & 45  & 65  \\

F & 23698 & 23823 & 20744 & 1362  & 24021 & 23977 & 24027 \\
  & 23837 & 23883 & 24004 & 23729 & 4736  & 3374  & 21313 \\
  & 12061 & 11014 & 16161 & 14877 & 16125 & 1297  & 1256  \\

G & 111535 & 111446 & 108299 & 65077 & 111831 & 111720 & 111841 \\
  & 111073 & 110958 & 111790 & 111701 & 74542  & 73388  & 111424 \\
  & 105506 & 92075  & 110991 & 98557  & 110372 & 53733  & 61161  \\

K & 42201 & 41759 & 41145 & 37818 & 42221 & 42193 & 42225 \\
  & 37039 & 40391 & 42221 & 42214 & 15124 & 15138 & 40977 \\
  & 37456 & 13079 & 42183 & 40887 & 40375 & 17804 & 12250 \\

M & 17305 & 17155 & 12815 & 17286 & 17304 & 17236 & 17306 \\
  & 3918  & 5305  & 17297 & 17299 & 1778  & 3897  & 1141  \\
  & 822   & 134   & 9073  & 17021 & 13523 & 3886  & 119   \\

\hline\hline
\end{tabular}
\label{tab:Valid Parameter Count_with_APOGEE}
\end{table*}

\section{Methodology} \label{sec:method}
    
\subsection{Sample Construction}

In this study, for the DR11 dataset labeled by the LAMOST pipeline, we applied the following filtering and preprocessing procedures to the raw data:

\begin{enumerate}
    \item \textbf{Signal-to-noise ratio thresholding}:  
    We first set a minimum signal-to-noise ratio (SNR) threshold of 2 in the $r$ band (SNRr) and the $i$ band (SNRi). Since the SNRs in the $u$, $g$, and $z$ bands are generally low, using these bands as filtering criteria would lead to substantial sample loss. Therefore, SNR control was performed only in the $r$ and $i$ bands.

    \item \textbf{Bad pixel masking}:  
    For pixels (wavelength points) with the \texttt{ormask} flag equal to 1, the corresponding flux values were masked (excluded from analysis). According to the LAMOST low-resolution data documentation, pixels with \texttt{ormask}=1 may suffer from one of the following issues: bad CCD pixels, abnormal profiles during spectral extraction, lack of sky information at that wavelength, excessively high sky background, fiber traces deviating from the CCD region, or the absence of valid data in that wavelength range.

    \item \textbf{Spectral normalization}:  
    To reduce sensitivity to flux calibration uncertainties (and overall continuum scaling), we applied the following normalization scheme (Eq.~\eqref{eq:flux-normalize}):
    \begin{equation}
        flux = \frac{flux}{\max(flux)},
        \label{eq:flux-normalize}
    \end{equation}

    \item \textbf{Spectral interpolation and resampling}:  
    To ensure consistent wavelength coverage and uniform wavelength spacing across all spectra, we resampled the spectra over the interval $[3800, 8700]\,\text{\AA}$ using cubic polynomial interpolation. Each spectrum was resampled with a step size of $1\,\text{\AA}$, resulting in a spectral vector of dimension 4901.

    \item \textbf{Flux value precision}:  
    All spectral flux values were rounded to three decimal places. Given the limited context length of large language models, retaining too many decimal places would reduce the number of pixels that can be input. After trade-off considerations, a precision of three decimal places was adopted.
\end{enumerate}
    
The QA-pair format for the two-stage training was designed with the following considerations:
\begin{enumerate}
    \item The normalized spectral flux values were inserted between \texttt{<flux>} and \texttt{</flux>}, ordered by wavelength from $\rm 3800\,\AA$ to $\rm 8700\,\AA$ with a step of $\rm 1\,\AA$. Each flux value was retained to three decimal places.
    
    \item To facilitate character-level tokenization of numerical strings by the large language model, we adopted the method proposed by \citet{gruver2023llmtime}, in which spaces are inserted between numbers and punctuation marks in the “answer” section to enable character-level tokenization.
    
    \item In the QA pairs constructed for continued training on LAMOST spectra cross-matched with APOGEE and annotated with APOGEE-derived parameter labels, an identical parameter set and the same number of parameter fields were adopted in the “question” and “answer” sections.
\end{enumerate}

The QA designs for the two training stages are summarized as follows:
\begin{itemize}
  \item {For the QA design of the first-stage model}: 
\begin{lstlisting}
{
  "question": "Given the normalized flux data sorted by wavelength, from 3800 angstrom to 8700. Your task is to analyze this spectrum, and predict spectral type and physical parameters. Organize your answer in the order of ['class', 'Teff', 'logg', '[Fe/H]']. <flux>0.303, 0.307, 0.302, 0.298, 0.3, 0.296, 0.295, 0.297, ...</flux>",
  "answer": "[ 'M',' 3 7 9 4 . 6 6 ',' 4 . 6 2 8 ',' - 0 . 0 2 9 5 ' ]"
}
\end{lstlisting}

  \item {For the QA design of the second-stage continued training model}:
\begin{lstlisting}
{
  "question": "Given the normalized flux data sorted by wavelength, from 3800 angstrom to 8700 angstrom. Your task is to analyze this spectrum, and predict physical parameters: C_FE CI_FE N_FE O_FE NA_FE MG_FE. <flux>0.303, 0.307, 0.302, 0.298, 0.3, 0.296, ...</flux>",
  "answer": "[C_FE:' 0 . 1 5 1 3 9 9 9 9 ', CI_FE:' 0 . 1 3 8 5 1 9 9 9 ', N_FE:' - 0 . 0 2 6 2 5 4 0 2 3 ', O_FE:' 0 . 2 2 6 4 5 4 9 9 ', NA_FE:' - 0 . 0 1 3 0 6 3 9 8 1 ', MG_FE:' 0 . 2 7 5 5 5 1 7 ']"
}
\end{lstlisting}
\end{itemize}

\subsection{Model Structure}

The model employed in this study is \textit{LLaMA-3.1-8B Instruct}, which is one of the LLaMA-3 series models released by Meta AI in 2024. This model contains approximately 8 billion parameters and adopts a decoder-only Transformer architecture, following the self-attention mechanism proposed by \cite{NIPS2017_3f5ee243}. It was pre-trained on multilingual corpora with trillions of tokens, enabling the model to acquire general-purpose sequence modeling capabilities. These capabilities include identifying both local and global dependencies (e.g., subject--verb agreement and long-range coreference in syntactic structures), parsing hierarchical compositional structures (e.g., nested clauses and tree-like semantic representations), and learning context-dependent dynamic representations. Fundamentally, the model abstracts generalizable sequence patterns from massive data, rather than being bound to surface forms of any specific language.

In our implementation, the input text is constructed by serializing the normalized spectral flux values (as numerical strings), allowing the model to treat stellar spectra as quasi-linguistic sequences. Through stage-wise training, the model is able to model long-range dependencies between absorption and emission features across the wavelength range of 3800--8700\,\AA, thereby extracting structured latent representations from the spectra—namely, the stellar physical states jointly encoded by spectral line distributions, strengths, and profiles. These learned spectral patterns are ultimately mapped to key atmospheric parameters, including the effective temperature (\textit{T$_\mathrm{eff}$}), surface gravity (\textit{log\,$g$}), and metallicity ([Fe/H]).

\subsection{Training Methods}

We first pre-trained the model using 1.5 million LAMOST spectra, with the primary objective of transferring the model’s sequence modeling capability to spectroscopic data. Subsequently, we fine-tuned the model on approximately $10^5$ LAMOST spectra annotated with APOGEE-derived elemental abundance labels, to enhance its performance on downstream multi-physical-parameter prediction tasks. During training, we employed QLoRA (Quantized Low-Rank Adaptation) to optimize resource utilization, significantly reducing GPU memory consumption while maintaining high computational precision. The rank was set to 64, the scaling factor to 128, and a dropout rate of 10\% was introduced to mitigate overfitting.

To further improve training efficiency, gradient accumulation and gradient checkpointing techniques were enabled, allowing for larger effective batch sizes and longer context lengths. Mixed-precision training was implemented using \texttt{bf16}, in conjunction with DeepSpeed’s ZeRO Stage~3 optimization strategy to ensure efficient memory management and scalable distributed training. The AdamW optimizer was used with an initial learning rate of 0.0001 and a weight decay coefficient of 0.1, while a cosine annealing scheduler was applied to dynamically adjust the learning rate. The dataset was split into training and testing subsets with a ratio of 8:2.

\section{Results} \label{sec:method on machine learning}

This section presents and analyzes the experimental results of the model on the test set as well as on external comparison datasets. During sample construction, for each element, only abundance measurements with a flag value of zero were included in the statistics.

\subsection{Test Results of the Single-Stage LLM Model} \label{sec:stage1_results}

To comprehensively evaluate the performance of the single-stage model, we first report its overall metrics on the test set, and then further compare the inference results of our model with those of other widely used models.

\paragraph{Overview of Test Set Results}

As shown in Table~\ref{tab:internal_errors}, the performance metrics of the model on the test split of the 1.5 million-sample dataset are presented. The results indicate that the model achieves low bias and dispersion in predicting the three atmospheric parameters across different stellar types.

\begin{table}[htbp]
\centering
\caption{Prediction residual statistics of the first-stage model on the $1.5\times10^6$-sample test set}
\begin{tabular}{lccc}
\hline
\hline
Subclass & $T_{\mathrm{eff}}$ ($\mu \pm \sigma$) & $\log g$ ($\mu \pm \sigma$) & [Fe/H] ($\mu \pm \sigma$) \\
\hline
A & $17 \pm 71$ & $0.01 \pm 0.06$ & $-0.02 \pm 0.05$ \\
F & $0 \pm 30$ & $0.01 \pm 0.05$ & $-0.02 \pm 0.02$ \\
G & $-5 \pm 35$ & $-0.02 \pm 0.07$ & $-0.02 \pm 0.03$ \\
K & $-1 \pm 32$ & $-0.01 \pm 0.06$ & $-0.02 \pm 0.04$ \\
M & $-1 \pm 39$ & $0.00 \pm 0.07$ & $-0.01 \pm 0.11$ \\
All Classes & $0 \pm 38$ & $0.00 \pm 0.06$ & $-0.02 \pm 0.04$ \\
\hline
\end{tabular}
\label{tab:internal_errors}
\end{table}

\paragraph{Comparison with Other Models on the Test Set}

In addition, we compare the proposed method with several traditional machine learning approaches, including the Light Gradient Boosting Machine (LightGBM), convolutional neural networks (CNNs), and recurrent neural networks (RNNs). The dataset used here consists of 100,000 spectra, covering five stellar spectral types (A, F, G, K, and M), with approximately 20,000 samples for each type. The full dataset was split with an 8:2 ratio, where 80\% was used for training and 20\% for testing. The comparative results of the different models on the test set are presented in Table~\ref{tab:ML_diff_vs_LLM}. The parameter settings and key training details for each model are summarized as follows:

\textbf{LightGBM}: LightGBM \citep{meng16lightGBM,ke17lightGBM} is a model based on Gradient Boosting Decision Trees (GBDTs). Compared with other GBDT-based models (e.g., XGBoost; \citealp{Chen2016xgboost}), LightGBM adopts a leaf-wise growth strategy rather than a level-wise one, significantly reducing training time while maintaining strong predictive performance. Combined with efficient parallel computing, LightGBM generally achieves faster training speeds than most GBDT-based models. In the astronomical domain, \citet{Liang_2022GBM} applied LightGBM to stellar parameter inversion using simulated photometric data from the China Space Station Telescope (CSST). Since LightGBM cannot simultaneously output classification and regression results, we trained separate models for classification and parameter regression in this study. The input consists of spectra, while the outputs are stellar spectral types for classification and $T_{\rm eff}$, $\log g$, and [Fe/H] for regression.

\textbf{CNN}: For the CNN-based model, we adopted ResNet-50 \citep{he16resnet} to evaluate its performance on our stellar catalog. By introducing residual connections, ResNet allows information and gradients to propagate directly from higher layers to lower layers, alleviating the degradation problem commonly encountered in deep CNN training and enabling deeper architectures to improve performance. For the spectral inputs in this study, the input layer of ResNet-50 was adapted to match the structure of the spectral data, and the output layer consisted of one classification head and three regression heads. The training loss was defined as the sum of the classification loss and 50 times the regression loss, reflecting our emphasis on the accuracy of stellar physical parameters over spectral type classification. The classification loss was computed using cross-entropy, while the regression loss was based on mean squared error (MSE). The initial learning rate was set to 0.001, the total number of training epochs to 80, and the mini-batch size to 32. If the validation loss did not decrease for three consecutive epochs, the learning rate was reduced by a factor of 0.1; if no improvement was observed for seven consecutive epochs, early stopping was applied. For additional comparison, a simpler CNN architecture with fewer parameters was also included (denoted as Small CNN).

\textbf{RNN}: For the RNN-based model, we selected the Gated Recurrent Unit (GRU; \citealp{cho2014gru}) as a representative architecture to evaluate the performance of recurrent neural networks on this stellar catalog task. Compared with Long Short-Term Memory networks (LSTM; \citealp{hochr1997lstm}), GRUs employ a single gating mechanism to control both information update and forgetting, achieving faster training while retaining strong modeling capacity \citep{rnet2017}. In astronomy, \citet{chaini2020grulc} applied GRU models to classify light curves from the PLAsTiCC dataset\footnote{\url{https://plasticc.org/}}. In this study, the input and output layers of the GRU model were adapted to the spectral data structure. The model consisted of three stacked GRU modules, and the loss function was identical to that used for ResNet-50. Training was stopped early at epoch 58, with a maximum of 80 epochs specified.

As shown in Table~\ref{tab:ML_diff_vs_LLM}, the different models exhibit clear differences in their predictive performance for stellar parameters, such as the effective temperature (\textit{T$_\mathrm{eff}$}), surface gravity (\textit{log\,$g$}), and metallicity ([Fe/H]). Overall, \textbf{our LLM model} achieves the smallest prediction errors across all spectral types (A, F, G, K, and M), outperforming the other models in the prediction of \textit{T$_\mathrm{eff}$}, \textit{log\,$g$}, and [Fe/H].

Specifically, the \textbf{LightGBM} model shows relatively stable performance, particularly for [Fe/H], where the errors are comparatively small. However, its errors in predicting effective temperature and surface gravity are larger; for example, for A-type stars, the performance in \textit{T$_\mathrm{eff}$} is $-13 \pm 433$~K.

In contrast, \textbf{ResNet-50} exhibits relatively strong overall performance, especially for \textit{log\,$g$} and [Fe/H], with smaller errors. In particular, for F-, G-, and K-type stars, most of the dispersions $\sigma$ are below 0.2~dex. However, the \textbf{Small CNN} model shows larger errors in predicting [Fe/H] and \textit{log\,$g$}, with most $\sigma$ values exceeding 0.2~dex, and also larger errors in \textit{T$_\mathrm{eff}$}, indicating overall lower predictive accuracy compared with ResNet-50.

The \textbf{GRU model} exhibits performance comparable to that of the Small CNN, with relatively large errors in \textit{T$_\mathrm{eff}$} and also larger errors in \textit{log\,$g$} and [Fe/H], where most $\sigma$ values exceed 0.2~dex.

Overall, the LLM yields consistently lower dispersion across spectral types, indicating improved stability for stellar-label inference on low-resolution spectra.

\begin{deluxetable}{c|ccc}
\tablecaption{Comparison of prediction residuals between traditional machine-learning models and our LLM on the $10^5$-sample test dataset \label{tab:ML_diff_vs_LLM}}
\tabletypesize{\footnotesize}
\tablewidth{0pt}

\tablehead{
\multicolumn{1}{c|}{Class} & \multicolumn{1}{c}{$T_{\rm eff}$ (K)} &
\multicolumn{1}{c}{log$g$ (dex)} & \multicolumn{1}{c}{[Fe/H] (dex)}
}

\startdata
\multicolumn{4}{c}{\textbf{LightGBM}} \\
\hline
A & $-13 \pm 433$ & $-0.01 \pm 0.19$ & $0.01 \pm 0.27$ \\
F & $-21 \pm 185$ & $0.02 \pm 0.21$  & $-0.01 \pm 0.19$ \\
G & $6 \pm 170$   & $0.01 \pm 0.34$  & $0.01 \pm 0.18$ \\
K & $-12 \pm 168$ & $-0.01 \pm 0.27$ & $0.01 \pm 0.19$ \\
M & $39 \pm 176$  & $-0.01 \pm 0.41$ & $-0.02 \pm 0.19$ \\
Overall & $-0 \pm 251$ & $0.00 \pm 0.30$ & $-0.00 \pm 0.21$ \\
\hline
\\[-6pt]
\multicolumn{4}{c}{\textbf{ResNet-50}} \\
\hline
A & $86 \pm 666$  & $0.01 \pm 0.18$ & $0.05 \pm 0.22$ \\
F & $-29 \pm 180$ & $0.09 \pm 0.18$ & $0.06 \pm 0.14$ \\
G & $-39 \pm 132$ & $0.14 \pm 0.24$ & $0.09 \pm 0.12$ \\
K & $-53 \pm 165$ & $0.02 \pm 0.17$ & $0.06 \pm 0.13$ \\
M & $44 \pm 706$  & $-0.04 \pm 0.38$ & $0.01 \pm 0.16$ \\
Overall & $2 \pm 453$ & $0.04 \pm 0.25$ & $0.05 \pm 0.16$ \\
\hline
\\[-6pt]
\multicolumn{4}{c}{\textbf{Small CNN}} \\
\hline
A & $-94 \pm 699$ & $0.05 \pm 0.27$  & $0.01 \pm 0.34$ \\
F & $-50 \pm 328$ & $-0.03 \pm 0.29$ & $-0.01 \pm 0.23$ \\
G & $-2 \pm 288$  & $-0.03 \pm 0.43$ & $-0.02 \pm 0.22$ \\
K & $12 \pm 302$  & $-0.03 \pm 0.38$ & $-0.00 \pm 0.22$ \\
M & $53 \pm 396$  & $-0.07 \pm 0.64$ & $-0.01 \pm 0.25$ \\
Overall & $-17 \pm 434$ & $-0.02 \pm 0.42$ & $-0.01 \pm 0.26$ \\
\hline
\\[-6pt]
\multicolumn{4}{c}{\textbf{GRU}} \\
\hline
A & $99 \pm 849$  & $-0.01 \pm 0.21$ & $0.03 \pm 0.40$ \\
F & $-7 \pm 252$  & $0.02 \pm 0.26$  & $-0.03 \pm 0.33$ \\
G & $-0 \pm 243$  & $0.02 \pm 0.46$  & $0.04 \pm 0.31$ \\
K & $-8 \pm 233$  & $0.04 \pm 0.33$  & $-0.01 \pm 0.28$ \\
M & $6 \pm 238$   & $0.02 \pm 0.36$  & $-0.01 \pm 0.19$ \\
Overall & $18 \pm 440$ & $0.02 \pm 0.34$ & $0.01 \pm 0.31$ \\
\hline
\\[-6pt]
\multicolumn{4}{c}{\textbf{our LLM}} \\
\hline
A & $51 \pm 297$ & $0.05 \pm 0.15$  & $-0.00 \pm 0.18$ \\
F & $9 \pm 80$   & $0.01 \pm 0.14$  & $-0.03 \pm 0.09$ \\
G & $4 \pm 84$   & $-0.00 \pm 0.17$ & $-0.02 \pm 0.08$ \\
K & $8 \pm 72$   & $-0.01 \pm 0.15$ & $-0.02 \pm 0.09$ \\
M & $-1 \pm 53$  & $0.01 \pm 0.22$  & $-0.00 \pm 0.15$ \\
Overall & $14 \pm 150$  & $0.01 \pm 0.17$  & $-0.01 \pm 0.12$ \\
\hline
\enddata
\tablecomments{This table compares the mean and standard deviation of the difference between the predictions and labels of the stellar parameters ($T_{\rm eff}$, log$g$, and [Fe/H]) for different classes of stars using LightGBM, ResNet-50, GRU, Small CNN, and LLM models. Column 'Class' stands for the type of stars, and the overall stands for the full test set. The values are shown in the format of $\mu \pm \sigma$ for each stellar parameter.}

\end{deluxetable}

To further evaluate the performance of CNN architectures on larger-scale datasets, we constructed an expanded dataset based on the LAMOST 11th data release (DR11), with a total of $10^5$ stellar spectra. The model adopts the ResNet-50 architecture. A cosine annealing learning rate schedule was employed, with the learning rate gradually decaying from 0.001 to 0 over 50 training epochs.

To compare the parameter regression performance of CNNs and large language models on large-scale samples, we further evaluated the LLM (hereafter referred to as \emph{our LLM}) on the same $10^5$-sample dataset under the same dataset split and evaluation protocol, and conducted a spectral-type–wise statistical comparison of its prediction errors with those of ResNet-50. Table~\ref{tab:resnet_llm_diff} reports the $\mu \pm \sigma$ of the differences between predictions and labels for each spectral class (A/F/G/K/M) as well as for the overall sample. Here, $\mu$ represents the systematic bias, while $\sigma$ denotes the dispersion (precision).

From the overall results, \emph{our LLM} exhibits substantially better stability across all three key stellar parameters. For $T_{\rm eff}$, the dispersion of ResNet-50 was $\sigma = 908\,\mathrm{K}$, whereas \emph{our LLM} reduced it markedly to $\sigma = 133\,\mathrm{K}$. For log$g$, the dispersion decreased from 1.20 to 0.15, and for [Fe/H], it dropped from 0.54 to 0.11. We noted that ResNet-50 showed larger $\mu$ and $\sigma$ values for M-type stars, which was likely related to the generally lower signal-to-noise ratios of M-type spectra compared to other spectral types (as described in the sample construction procedure in Section~3.1, where a minimum SNR of 2 was adopted). This suggested that, when predicting multiple spectral parameters simultaneously, the LLM remained more stable and reliable than the deep learning model ResNet-50, particularly for low-SNR samples.

A spectral-type–resolved comparison further shows that, for $T_{\rm eff}$, log$g$, and [Fe/H], \emph{our LLM} systematically achieves smaller error dispersions for F-, G-, K-, and M-type stars. For the M-type subsample, the error distributions of ResNet-50 are extremely dispersed ($T_{\rm eff}$: $\sigma = 1975\,\mathrm{K}$; log$g$: $\sigma = 2.65$; [Fe/H]: $\sigma = 1.16$), whereas \emph{our LLM} maintains reasonable and stable error levels within the same class ($T_{\rm eff}$: $\sigma = 56\,\mathrm{K}$; log$g$: $\sigma = 0.23$; [Fe/H]: $\sigma = 0.14$). This supports the improved robustness of LLMs in subsamples with low signal-to-noise ratios.

It should be noted that for A-type stars, both models exhibit relatively large dispersions in $T_{\rm eff}$ (ResNet-50: $327\,\mathrm{K}$; \emph{our LLM}: $269\,\mathrm{K}$), indicating that early-type stellar parameter regression remains a shared challenge under the current data and modeling conditions. Nevertheless, even in this case, \emph{our LLM} still achieves smaller overall systematic biases and dispersions than ResNet-50 for A-type stars. Overall, the results presented in Table~\ref{tab:resnet_llm_diff} demonstrate that on a large dataset of 500k samples, \emph{our LLM} can markedly reduce the prediction dispersions of $T_{\rm eff}$, log$g$, and [Fe/H] compared with ResNet-50, while simultaneously achieving smaller systematic biases and more stable error distributions across most spectral types.

\begin{deluxetable}{c|ccc}
\tablecaption{Comparison of prediction residuals between ResNet-50 and our LLM on the $5\times10^5$-sample test dataset \label{tab:resnet_llm_diff}}
\tabletypesize{\footnotesize}
\tablewidth{0pt}

\tablehead{
\multicolumn{1}{c|}{Class} &
\multicolumn{1}{c}{$T_{\rm eff}$ (K)} &
\multicolumn{1}{c}{log$g$ (dex)} &
\multicolumn{1}{c}{[Fe/H] (dex)}
}

\startdata
\multicolumn{4}{c}{\textbf{ResNet-50}} \\
\hline
A       & $ 61 \pm 327$  & $ 0.05 \pm 0.14$ & $ 0.01 \pm 0.19$ \\
F       & $-47 \pm 137$  & $-0.08 \pm 0.14$ & $-0.03 \pm 0.12$ \\
G       & $-48 \pm 124$  & $-0.11 \pm 0.19$ & $-0.02 \pm 0.16$ \\
K       & $  2 \pm  93$  & $-0.09 \pm 0.15$ & $-0.02 \pm 0.12$ \\
M       & $107 \pm 1975$ & $-0.09 \pm 2.65$ & $-0.08 \pm 1.16$ \\
Overall & $ 15 \pm 908$  & $-0.06 \pm 1.20$ & $-0.03 \pm 0.54$ \\
\hline
\\[-6pt]
\multicolumn{4}{c}{\textbf{our LLM}} \\
\hline
A       & $22 \pm 269$ & $ 0.01 \pm 0.12$ & $-0.03 \pm 0.15$ \\
F       & $ 0 \pm  68$ & $ 0.00 \pm 0.12$ & $-0.03 \pm 0.07$ \\
G       & $-5 \pm  66$ & $-0.03 \pm 0.14$ & $-0.03 \pm 0.06$ \\
K       & $-4 \pm  55$ & $-0.02 \pm 0.11$ & $-0.03 \pm 0.08$ \\
M       & $-1 \pm  56$ & $ 0.01 \pm 0.23$ & $-0.01 \pm 0.14$ \\
Overall & $ 2 \pm 133$ & $0.00 \pm 0.15$ & $-0.03 \pm 0.11$ \\
\hline
\enddata

\tablecomments{The values are shown as $\mu \pm \sigma$ of the differences between predictions and labels for each stellar parameter.}
\end{deluxetable}

\subsection{Test Results of the Second-Stage \textit{LLM} Model} \label{sec:stage2_results}

To comprehensively evaluate the performance of the second-stage (continued-trained) model, we first present its overall performance metrics on the test set, and then further compare the model outputs with the parameter predictions obtained from ResNet-50.

\paragraph{Overview of Test Set Results}
The test set results are summarized in Table~\ref{tab:mu_sigma_stage_2_model}, which reports the prediction error statistics of the second-stage model for multiple elemental abundance ratios ($[\mathrm{X/Fe}]$) across different stellar spectral types (A/F/G/K/M). Here, the mean $\mu$ represents the systematic bias, while the dispersion $\sigma$ quantifies the scatter (precision) of the prediction errors. The sample counts shown in parentheses indicate that the distribution of spectral types in the test set is highly imbalanced: G-type and K-type stars dominate the sample, followed by F-type stars, whereas A-type and M-type stars are relatively scarce.

Overall, the mean biases $\mu$ for most elements across different spectral types are close to zero, indicating that the model does not show large systematic offsets on the test set.

From the perspective of dispersion, the prediction errors for most elements in G-type and K-type stars satisfy $\sigma < 0.2$~dex, suggesting that the model can achieve relatively stable abundance estimates for these elements in well-sampled spectral types. For F-type stars, the dispersions of these elements increase to some extent, but the majority still remain below $\sigma < 0.2$~dex.

In contrast, several element abundances, such as \texttt{NA\_FE}, \texttt{P\_FE}, \texttt{CO\_FE}, \texttt{CE\_FE}, and \texttt{TIII\_FE}, exhibit $\sigma > 0.2$~dex across multiple spectral types, with this behavior being particularly pronounced for A-type stars. This suggests the relatively larger uncertainties associated with the abundance predictions of these elements.

From an overall comparison across spectral types, G-type and K-type stars show the best stability in elemental abundance predictions, whereas the fraction of elements with $\sigma > 0.2$~dex increases markedly for A-type and M-type stars. Nevertheless, the predicted means for all spectral types remain close to zero, indicating that the model errors are dominated by random scatter rather than systematic bias.

\begin{table*}[ht!]
\centering
\caption{Prediction residual statistics of elemental abundance parameters for the second-stage model across spectral types on the test set: mean $\mu$ and dispersion $\sigma$}
\label{tab:mu_sigma_stage_2_model}
\renewcommand{\arraystretch}{1.15}
\begin{tabular}{lccccc}
\hline\hline
\textbf{Parameter} & \textbf{A (1087)} & \textbf{F (5092)} & \textbf{G (22780)} & \textbf{K (8765)} & \textbf{M (1255)} \\
\hline
C\_FE    & 0.02$\pm$0.11 & 0.01$\pm$0.09 & -0.01$\pm$0.07 & -0.01$\pm$0.05 & -0.01$\pm$0.06 \\
CI\_FE   & 0.02$\pm$0.11 & 0.01$\pm$0.10 & 0.00$\pm$0.09 & 0.00$\pm$0.09 & -0.01$\pm$0.08 \\
N\_FE    & -0.04$\pm$0.24 & -0.01$\pm$0.20 & 0.00$\pm$0.11 & -0.01$\pm$0.12 & -0.01$\pm$0.14 \\
O\_FE    & 0.01$\pm$0.09 & 0.01$\pm$0.07 & 0.00$\pm$0.07 & -0.02$\pm$0.12 & -0.02$\pm$0.13 \\
MG\_FE   & 0.02$\pm$0.10 & 0.01$\pm$0.08 & 0.00$\pm$0.06 & -0.01$\pm$0.08 & -0.03$\pm$0.09 \\
AL\_FE   & 0.01$\pm$0.14 & 0.01$\pm$0.11 & 0.00$\pm$0.10 & -0.01$\pm$0.11 & -0.04$\pm$0.14 \\
SI\_FE   & 0.01$\pm$0.09 & 0.01$\pm$0.07 & 0.01$\pm$0.05 & 0.00$\pm$0.06 & -0.01$\pm$0.06 \\
S\_FE    & -0.01$\pm$0.14 & 0.00$\pm$0.12 & 0.00$\pm$0.12 & 0.00$\pm$0.12 & -0.01$\pm$0.15 \\
MN\_FE   & 0.03$\pm$0.16 & 0.01$\pm$0.11 & 0.00$\pm$0.08 & -0.01$\pm$0.09 & 0.01$\pm$0.09 \\
NI\_FE   & 0.00$\pm$0.23 & 0.00$\pm$0.20 & 0.00$\pm$0.15 & -0.01$\pm$0.11 & -0.01$\pm$0.07 \\
CA\_FE   & 0.01$\pm$0.15 & 0.00$\pm$0.11 & 0.00$\pm$0.06 & -0.01$\pm$0.06 & -0.02$\pm$0.08 \\
NA\_FE   & 0.06$\pm$0.61 & 0.03$\pm$0.29 & 0.03$\pm$0.26 & 0.00$\pm$0.18 & 0.00$\pm$0.17 \\
P\_FE    & 0.01$\pm$0.49 & 0.03$\pm$0.36 & 0.03$\pm$0.31 & 0.00$\pm$0.22 & -0.01$\pm$0.21 \\
K\_FE    & 0.03$\pm$0.20 & 0.02$\pm$0.18 & 0.00$\pm$0.12 & 0.00$\pm$0.12 & -0.02$\pm$0.13 \\
TI\_FE   & -0.01$\pm$0.21 & -0.01$\pm$0.15 & -0.01$\pm$0.11 & -0.02$\pm$0.08 & -0.03$\pm$0.11 \\
TIII\_FE & 0.03$\pm$0.26 & 0.02$\pm$0.23 & 0.02$\pm$0.21 & 0.02$\pm$0.26 & 0.02$\pm$0.26 \\
CR\_FE   & 0.00$\pm$0.29 & 0.00$\pm$0.24 & 0.00$\pm$0.17 & 0.00$\pm$0.11 & -0.01$\pm$0.10 \\
CO\_FE   & 0.11$\pm$0.55 & 0.07$\pm$0.43 & 0.02$\pm$0.35 & -0.01$\pm$0.38 & 0.01$\pm$0.26 \\
CU\_FE   & -0.02$\pm$0.34 & -0.01$\pm$0.28 & 0.00$\pm$0.21 & 0.01$\pm$0.15 & -0.02$\pm$0.15 \\
V\_FE    & 0.00$\pm$0.21 & -0.01$\pm$0.21 & -0.01$\pm$0.18 & 0.00$\pm$0.18 & 0.01$\pm$0.17 \\
CE\_FE   & 0.03$\pm$0.30 & 0.01$\pm$0.28 & 0.02$\pm$0.28 & 0.04$\pm$0.24 & 0.03$\pm$0.23 \\
\hline\hline
\end{tabular}

\vspace{0.5em}
{\footnotesize \textit{Note}—The number in parentheses indicates the number of samples in each class.}
\end{table*}

\begin{deluxetable*}{l|cccccc|cccccc}
\tablecaption{Test-set RMSE comparison of elemental abundance predictions between ResNet-50 and the second-stage LLM model across spectral types
\label{tab:rmse_model_comparison_stage_2}}
\tabletypesize{\scriptsize}
\tablewidth{0pt}
\tablehead{
Parameter &
\multicolumn{6}{c|}{ResNet-50} &
\multicolumn{6}{c}{LLM} \\
& All & A & F & G & K & M
& All & A & F & G & K & M
}
\startdata
\hline
C\_FE
& 0.08 & 0.11 & 0.08 & 0.09 & 0.06 & 0.06
& 0.11 & 0.23 & 0.12 & 0.10 & 0.10 & 0.12 \\
CI\_FE
& 0.11 & 0.11 & 0.09 & 0.13 & 0.08 & 0.08
& 0.14 & 0.15 & 0.13 & 0.14 & 0.14 & 0.15 \\
N\_FE
& 0.12 & 0.15 & 0.18 & 0.11 & 0.08 & 0.12
& 0.16 & 0.37 & 0.17 & 0.15 & 0.16 & 0.18 \\
MG\_FE
& 0.08 & 0.12 & 0.09 & 0.08 & 0.07 & 0.08
& 0.12 & 0.17 & 0.12 & 0.12 & 0.12 & 0.14 \\
AL\_FE
& 0.17 & 0.30 & 0.19 & 0.19 & 0.08 & 0.12
& 0.13 & 0.13 & 0.13 & 0.13 & 0.13 & 0.14 \\
SI\_FE
& 0.08 & 0.11 & 0.09 & 0.09 & 0.05 & 0.07
& 0.09 & 0.13 & 0.09 & 0.09 & 0.09 & 0.11 \\
S\_FE
& 0.18 & 0.12 & 0.11 & 0.21 & 0.13 & 0.07
& 0.16 & 0.18 & 0.16 & 0.16 & 0.16 & 0.17 \\
MN\_FE
& 0.10 & 0.12 & 0.11 & 0.11 & 0.06 & 0.06
& 0.10 & 0.14 & 0.10 & 0.10 & 0.10 & 0.11 \\
NI\_FE
& 0.08 & 0.14 & 0.09 & 0.08 & 0.05 & 0.08
& 0.06 & 0.12 & 0.06 & 0.06 & 0.06 & 0.07 \\
CA\_FE
& 0.14 & 0.16 & 0.11 & 0.18 & 0.04 & 0.06
& 0.09 & 0.08 & 0.09 & 0.09 & 0.09 & 0.09 \\
NA\_FE
& 0.21 & 0.26 & 0.30 & 0.21 & 0.13 & 0.10
& 0.27 & 0.24 & 0.27 & 0.27 & 0.27 & 0.28 \\
P\_FE
& 0.24 & 0.16 & 0.20 & 0.28 & 0.13 & 0.09
& 0.31 & 0.27 & 0.31 & 0.31 & 0.31 & 0.33 \\
K\_FE
& 0.47 & 0.28 & 0.24 & 0.59 & 0.11 & 0.07
& 0.16 & 0.36 & 0.16 & 0.15 & 0.15 & 0.17 \\
TI\_FE
& 0.28 & 0.18 & 0.16 & 0.34 & 0.08 & 0.05
& 0.14 & 0.09 & 0.14 & 0.14 & 0.14 & 0.14 \\
TIII\_FE
& 0.17 & 0.08 & 0.18 & 0.19 & 0.12 & 0.05
& 0.26 & 0.15 & 0.25 & 0.26 & 0.26 & 0.28 \\
CR\_FE
& 0.21 & 0.15 & 0.22 & 0.24 & 0.10 & 0.08
& 0.18 & 0.15 & 0.18 & 0.18 & 0.17 & 0.18 \\
CO\_FE
& 0.55 & 0.09 & 0.38 & 0.66 & 0.30 & 0.22
& 0.35 & 0.12 & 0.35 & 0.35 & 0.35 & 0.36 \\
CU\_FE
& 0.22 & 0.17 & 0.24 & 0.25 & 0.12 & 0.10
& 0.22 & 0.31 & 0.22 & 0.21 & 0.22 & 0.24 \\
O\_FE
& 0.08 & 0.10 & 0.07 & 0.08 & 0.06 & 0.07
& 0.13 & 0.11 & 0.14 & 0.13 & 0.13 & 0.14 \\
V\_FE
& 0.14 & 0.02 & 0.06 & 0.16 & 0.11 & 0.06
& 0.20 & 0.21 & 0.21 & 0.20 & 0.21 & 0.20 \\
CE\_FE
& 0.19 & 0.25 & 0.19 & 0.21 & 0.14 & 0.04
& 0.31 & 0.31 & 0.30 & 0.31 & 0.31 & 0.26 \\
\enddata
\tablecomments{
RMSE values for stellar parameters.
Columns A, F, G, K, and M correspond to spectral types.
}
\end{deluxetable*}

\paragraph{Comparison with Other Models on the Test Set}

To further evaluate the second-stage LLM model, we compare it with the traditional residual neural network baseline, ResNet-50, under the same data splits and evaluation metrics. Table~\ref{tab:rmse_model_comparison_stage_2} reports the root-mean-square error (RMSE) for different spectral-type subsamples (A/F/G/K/M) as well as for the full sample (All).

The dataset was split into training and testing sets with an 8:2 ratio. For ResNet-50, missing parameters were handled using masking to avoid their influence on the training process. During training, all parameters were normalized using min--max normalization.

The training loss was defined as the sum of the classification loss and the average mean squared error (MSE) over all stellar parameters, with the stellar-parameter loss weighted by a factor of 50, reflecting our emphasis on accurate prediction of stellar physical parameters over spectral-type classification. The initial learning rate was set to 0.001. When the validation loss failed to decrease for three consecutive epochs, the learning rate was reduced by a factor of two. Training was conducted for a total of 100 epochs without early stopping, and the model with the lowest validation loss was saved. The mini-batch size was set to 256.

 For several elements, the second-stage LLM achieves lower RMSEs than ResNet-50, indicating superior predictive accuracy. These elements include \texttt{AL\_FE} (0.13 vs.\ 0.17), \texttt{S\_FE} (0.16 vs.\ 0.18), \texttt{NI\_FE} (0.06 vs.\ 0.08), \texttt{CA\_FE} (0.09 vs.\ 0.14), \texttt{K\_FE} (0.16 vs.\ 0.47), \texttt{TI\_FE} (0.14 vs.\ 0.28), and \texttt{CR\_FE} (0.18 vs.\ 0.21). Among these, the improvements for \texttt{K\_FE} and \texttt{TI\_FE} are particularly pronounced, with RMSE reductions exceeding 50\%, highlighting a substantial improvement of the second-stage LLM for certain elements.

On the other hand, ResNet-50 achieves slightly lower RMSEs for a number of elemental abundance parameters, including \texttt{C\_FE} (0.08 vs.\ 0.11), \texttt{CI\_FE} (0.11 vs.\ 0.14), \texttt{N\_FE} (0.12 vs.\ 0.16), \texttt{MG\_FE} (0.08 vs.\ 0.12), \texttt{SI\_FE} (0.08 vs.\ 0.09), \texttt{NA\_FE} (0.21 vs.\ 0.27), \texttt{P\_FE} (0.24 vs.\ 0.31), \texttt{O\_FE} (0.08 vs.\ 0.13), \texttt{V\_FE} (0.14 vs.\ 0.20), \texttt{CE\_FE} (0.19 vs.\ 0.31), and \texttt{TIII\_FE} (0.17 vs.\ 0.26), suggesting that CNNs perform better for these abundances.

For some elements, the overall performance of the two models is comparable, such as \texttt{MN\_FE} (0.10 vs.\ 0.10) and \texttt{CU\_FE} (0.22 vs.\ 0.22). For \texttt{CO\_FE} (0.55 vs.\ 0.35), both models perform poorly, with large dispersions, suggesting that this element remains challenging under the current data/labeling conditions. From the spectral-type–resolved results, the dispersions of predictions for both models are slightly larger for A- and M-type stars than for F/G/K-type stars, likely influenced by the smaller sample sizes.

\section{Discussion} \label{sec:method on machine learning}
    \subsection{Spectra as Language}
\label{sec:baseline}

When applied to spectroscopic tasks, large language models (LLMs) show notable advantages over traditional methods. These advantages are evident not only on small-scale datasets but become even more prominent in large-scale scenarios, supporting the joint inference of multiple physical parameters across the spectroscopic analysis pipeline, and offering an accurate and scalable alternative for astronomical spectral data analysis. From the perspective of core performance metrics, LLMs exhibit high accuracy and stability in predicting key stellar physical parameters, particularly for fundamental quantities such as the effective temperature (\(T_{\rm eff}\)), surface gravity (\(\log g\)), and metallicity (\([\rm Fe/H]\))—outperforming conventional models. In large-scale tests with datasets of the order of \(5\times10^{5}\) samples, the performance gains of LLMs were especially striking: the dispersion in \(T_{\rm eff}\) predictions was substantially reduced from approximately 908~K for traditional models to 133~K, the dispersion in \(\log g\) decreases from 1.20 to 0.15, and the dispersion in \([\rm Fe/H]\) drops from 0.54 to 0.11. Improvements of this magnitude were observed consistently across most spectral types.

We also found improved robustness in parameter estimation from low S/N spectra when using the LLM. For stellar types with complex spectra, low signal-to-noise ratios, or highly imbalanced sample distributions (e.g., M-type stars), they are still able to maintain reasonable and stable predictive performance. For M-type stars, the dispersion in \(T_{\rm eff}\) predictions is only 56~K with the LLM, whereas the ResNet-50 baseline reach dispersions as high as 1975~K, highlighting a substantial performance difference. Similar advantages are observed for the dominant spectral types F, G, and K. Taking G-type stars as an example, the LLM achieves a \(T_{\rm eff}\) dispersion of only 66~K and a \([\rm Fe/H]\) dispersion as low as 0.06~dex, significantly outperforming the corresponding values of 124~K and 0.16~dex from traditional models. For K-type stars, the LLM also achieves superior control of dispersion, with values of 55~K for \(T_{\rm eff}\) and 0.08~dex for \([\rm Fe/H]\), both better than those of conventional approaches. In the prediction of multiple elemental abundances, the LLM performs on par with the traditional ResNet-50 model for most elements, while showing notably improved performance for specific elements (e.g., \texttt{K\_FE} and \texttt{TI\_FE}).

We observed a scaling trend in the performance of LLMs for spectroscopic tasks, whereby model behavior varied systematically with model capacity and training data volume. By comparing results obtained from datasets of \(10^{5}\), \(5\times10^{5}\), and \(1.5\times10^{6}\) samples (as shown in Tables~\ref{tab:ML_diff_vs_LLM}, \ref{tab:resnet_llm_diff}, and \ref{tab:internal_errors}), we found that, as the training sample size and effective model capacity increased together, the mean prediction biases for the three fundamental atmospheric parameters tended to move closer to zero, while the prediction dispersions showed stable convergence trends. Even at the \(10^{5}\)-sample level, the model already maintained nearly unbiased estimates across different spectral types; at the \(5\times10^{5}\) and \(1.5\times10^{6}\) scales, the standard deviations of \(T_{\rm eff}\), \(\log g\), and \([\rm Fe/H]\) were further compressed, and more consistent statistical properties were achieved across the major spectral classes.

This continuously improving performance with increasing scale is consistent with the scaling-law behavior revealed by large language models in natural language processing tasks, whereby growth in model parameter count and effective training data systematically enhances the representation of complex high-dimensional structures. In the context of spectral modeling—a task characterized by high dimensionality, strong correlations, and implicit physical constraints—the scaling behavior of LLMs is particularly critical. Larger model capacity may help improve the modeling accuracy of global spectral shapes and cross-band contextual relationships, and is likely to facilitate a more unified representation of physical distribution differences among spectral types, thereby contributing to improved predictive stability and generalization as the sample size increases. These results demonstrate that the paradigm of modeling spectra as a form of “language” inherently possesses strong scalability, providing a promising direction for further improvements in parameter inversion accuracy under larger samples, higher spectral resolution, and more complex observational conditions.

The superior performance of LLMs may be attributed to their unique mechanisms for global information integration and cross-contextual relationship mining. Compared with the CNN baseline used here (ResNet-50), which are often limited to local feature extraction and struggle to capture global correlations within spectral data, LLMs can deeply exploit the overall spectral structure and cross-band contextual information, can capture the complex nonlinear relationships between spectral morphology and stellar physical parameters. This mode of information processing is highly compatible with the intrinsically high-dimensional and strongly correlated nature of spectroscopic data, enabling LLMs to more fully learn the underlying physical laws from large-scale samples. As a result, they can mitigate common issues such as underfitting or convergence to local optima that often affect traditional models, leading to improved accuracy and robustness.

\subsection{Consistency of Inferred Parameters with External Survey Results}

In the previous subsection, we systematically demonstrated the validity and advantages of treating spectra as “language” and introducing large language models for spectroscopic modeling, from both the perspective of the performance gains observed in comparative experiments and the underlying reasons for these advantages. Validation on large-sample test sets shows that LLMs not only achieve significantly higher accuracy and stability than traditional models in predicting atmospheric physical parameters, but also deliver performance in multi-element abundance prediction tasks that is comparable to conventional neural network models, with clear advantages for certain elements. These results suggest the potential of LLMs to capture complex relationships between the global spectral structure and stellar physical parameters within the APOGEE spectral data framework.

Nevertheless, internal evaluations based solely on homogeneous datasets are insufficient to fully establish the scientific reliability and generalization capability of the model. To further assess the applicability of LLMs in realistic astronomical observational scenarios, it is necessary to cross-validate their inference results against external high-resolution spectroscopic surveys. Therefore, in the following analysis, we systematically compare the model predictions with data from external high-resolution surveys such as Gaia-ESO and GALAH DR4, in order to evaluate the consistency of inference and the physical plausibility of the model under cross-dataset conditions.

\subsubsection{Comparison between the First-Stage Model and External Datasets}

To examine the consistency and generalization capability of the model on external high-resolution spectroscopic datasets, we compared the inference results of the single-stage LLM model with stellar parameters provided by \textbf{Gaia-ESO} and \textbf{GALAH DR4}, respectively. We focused on three key parameters—$T_{\mathrm{eff}}$, $\log g$, and [Fe/H]—and evaluated their systematic biases and dispersions across different spectral-type subsamples. The corresponding statistical results are summarized in Table~\ref{tab:Gaia_ESO_galah_dr4_stage_1_model}. For the GALAH dataset, the adopted data quality cuts\footnote{\url{https://www.galah-survey.org/dr4/using_the_data/}} were
\texttt{snr\_lpx\_ccd3 > 30 \&\& flag\_sp == 0 \&\& flag\_X\_fe == 0} \citep{2025PASA...42...51B}.
Due to the limited number of stars cross-matched with Gaia-ESO, no additional quality cuts were applied to that dataset.

\begin{figure*}[ht!]
\centering
\includegraphics[width=\textwidth, trim=0cm 0cm 0cm 0cm, clip]{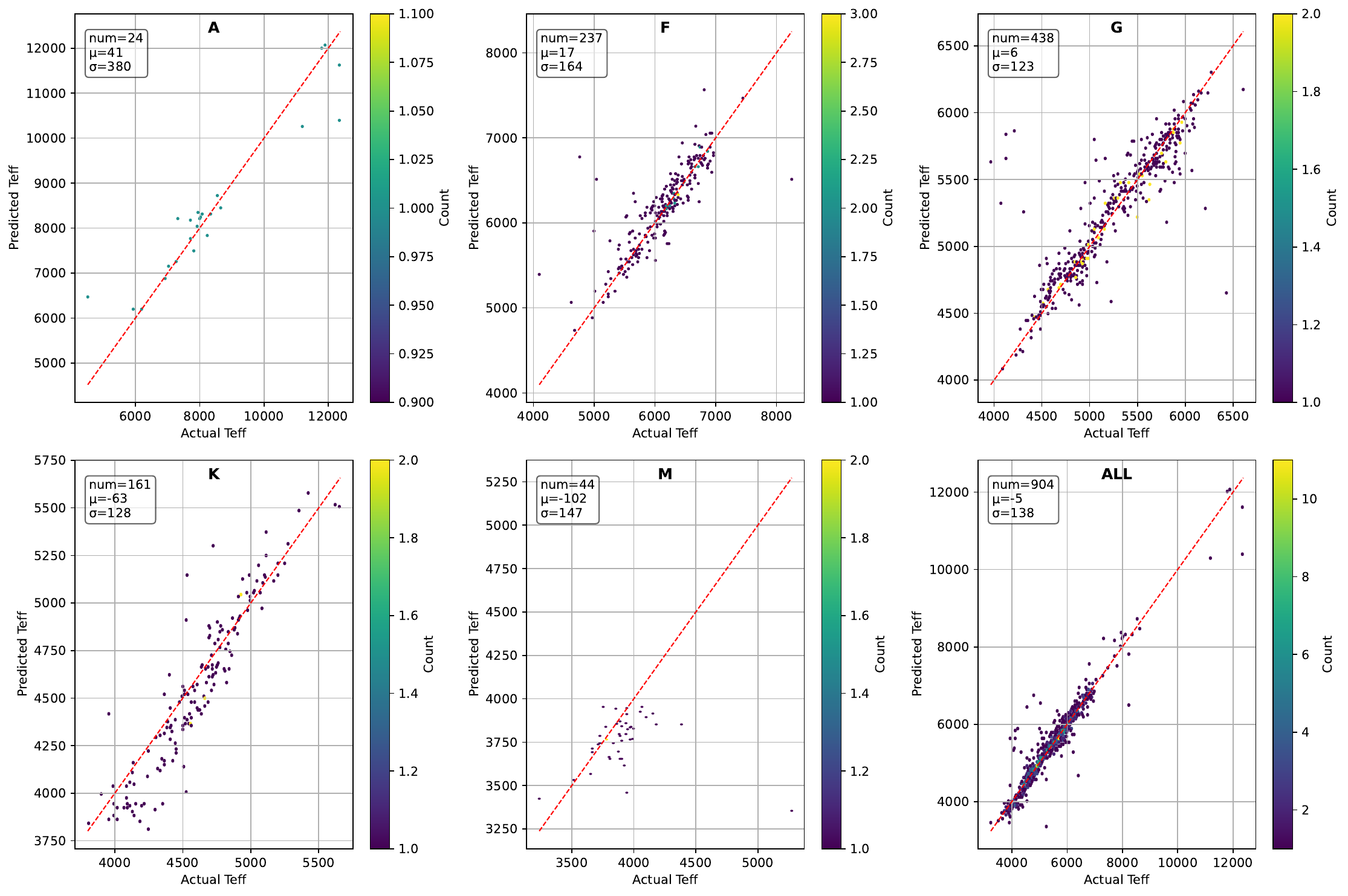}
\caption{Teff compared with Gaia-ESO.}
\label{fig:Teff_Gaia-ESO}
\end{figure*}

\begin{figure*}[ht!]
\centering
\includegraphics[width=\textwidth, trim=0cm 0cm 0cm 0cm, clip]{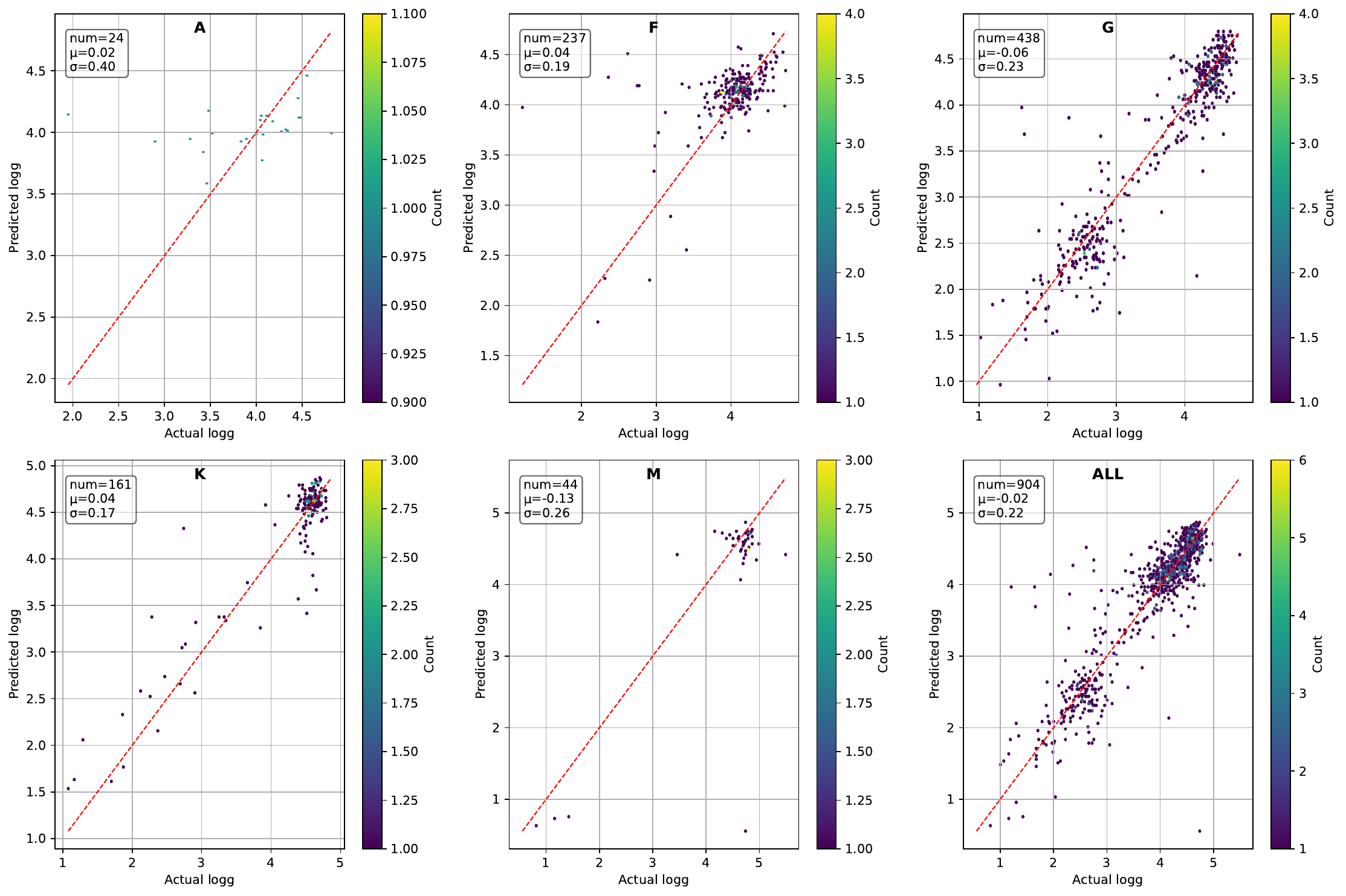}
\caption{logg compared with Gaia-ESO.}
\label{fig:logg_Gaia-ESO}
\end{figure*}

\begin{figure*}[ht!]
\centering
\includegraphics[width=\textwidth, trim=0cm 0cm 0cm 0cm, clip]{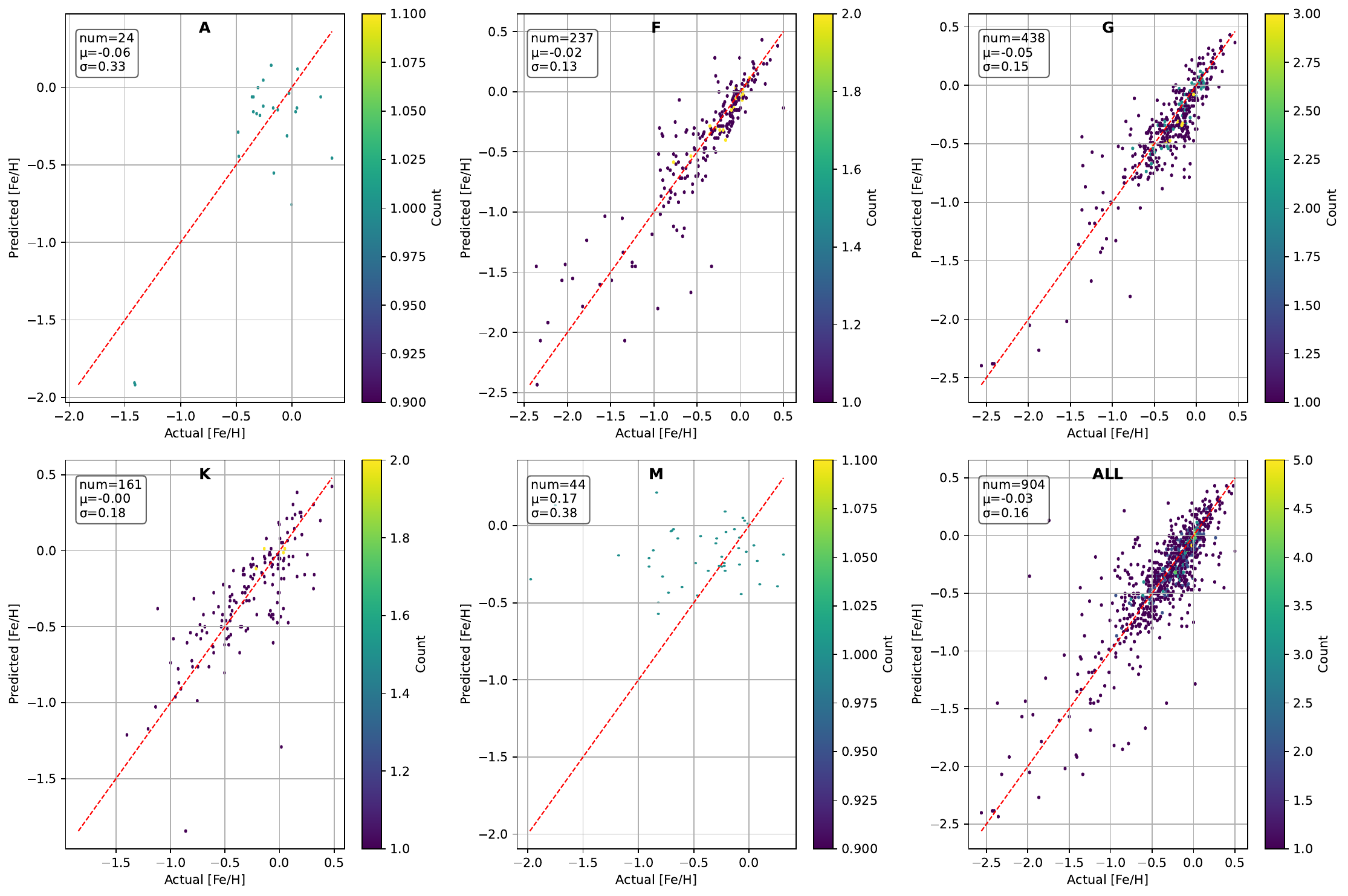}
\caption{[Fe/H] compared with Gaia-ESO.}
\label{fig:FeH_Gaia-ESO}
\end{figure*}

For the \textbf{Gaia-ESO} dataset, Figs.~\ref{fig:Teff_Gaia-ESO}, \ref{fig:logg_Gaia-ESO}, and \ref{fig:FeH_Gaia-ESO} illustrate the comparisons between the model predictions and the reference labels. As shown in Table~\ref{tab:Gaia_ESO_galah_dr4_stage_1_model}, for F-, G-, and K-type stars, the mean biases of all three parameters were generally small—mostly below 100~K—and the dispersions $\sigma$, although somewhat larger, remained close to the $10^2$~K level. This suggested good consistency between the model predictions and the Gaia-ESO scale for these well-sampled spectral types. In contrast, A- and M-type stars exhibited relatively larger dispersions in $T_{\mathrm{eff}}$, $\log g$, and [Fe/H], which were likely influenced by their limited sample sizes and the resulting statistical uncertainties.

\begin{figure*}[ht!]
\centering
\includegraphics[width=\textwidth, trim=0cm 0cm 0cm 0cm, clip]{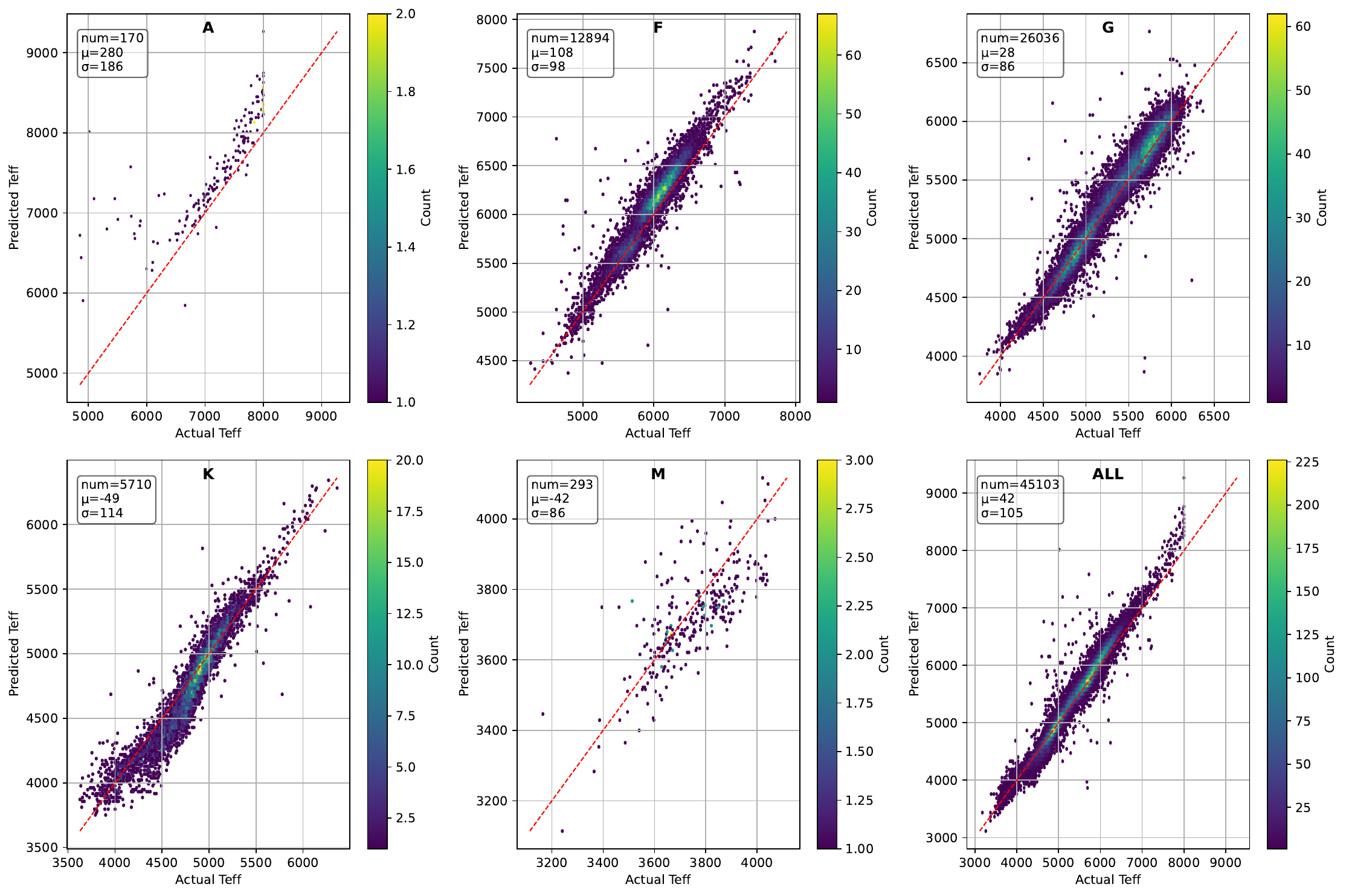}
\caption{Teff compared with Galah DR4.}
\label{fig:Teff_Galah-DR4}
\end{figure*}

\begin{figure*}[ht!]
\centering
\includegraphics[width=\textwidth, trim=0cm 0cm 0cm 0cm, clip]{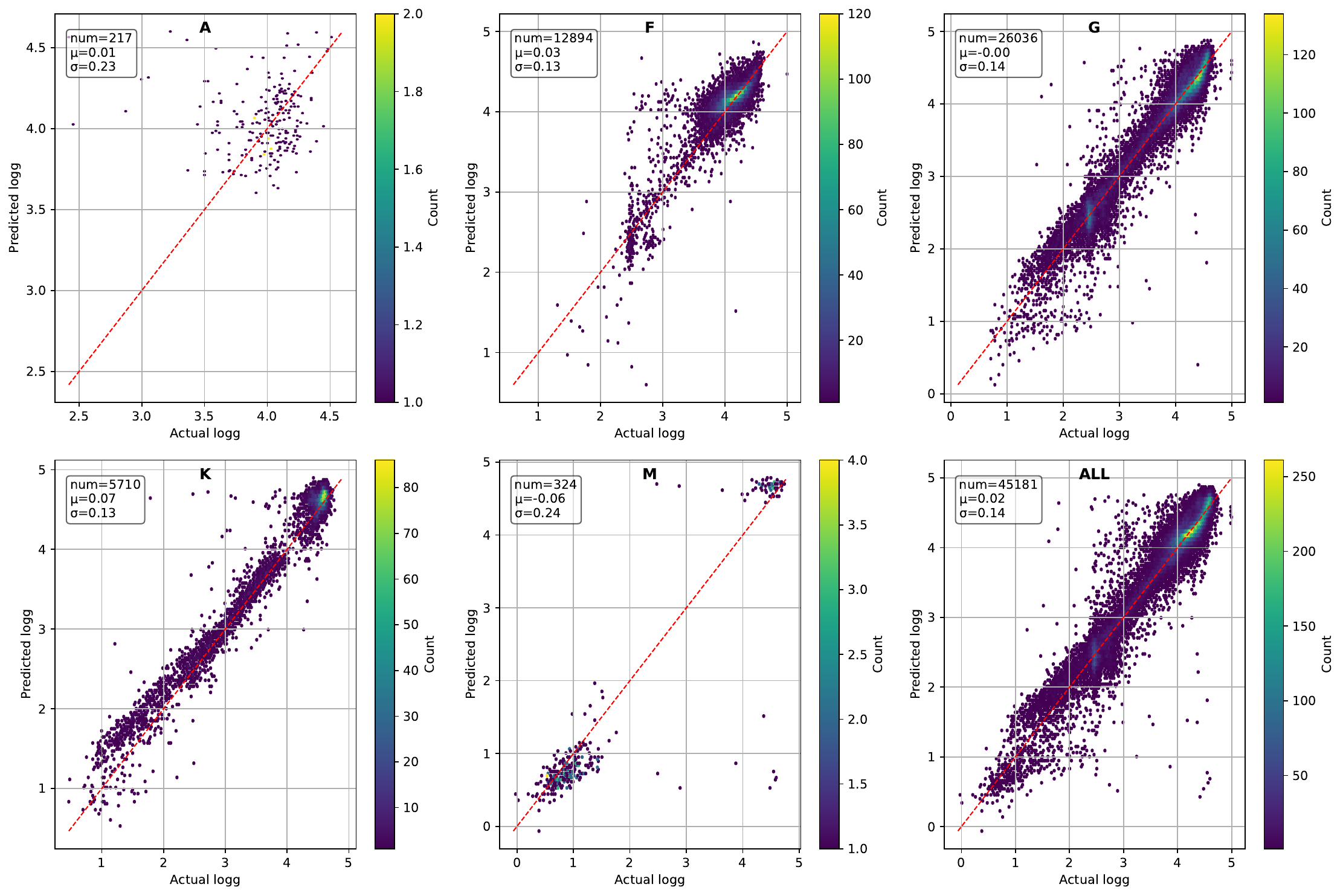}
\caption{logg compared with Galah DR4.}
\label{fig:logg_Galah-DR4}
\end{figure*}

\begin{figure*}[ht!]
\centering
\includegraphics[width=\textwidth, trim=0cm 0cm 0cm 0cm, clip]{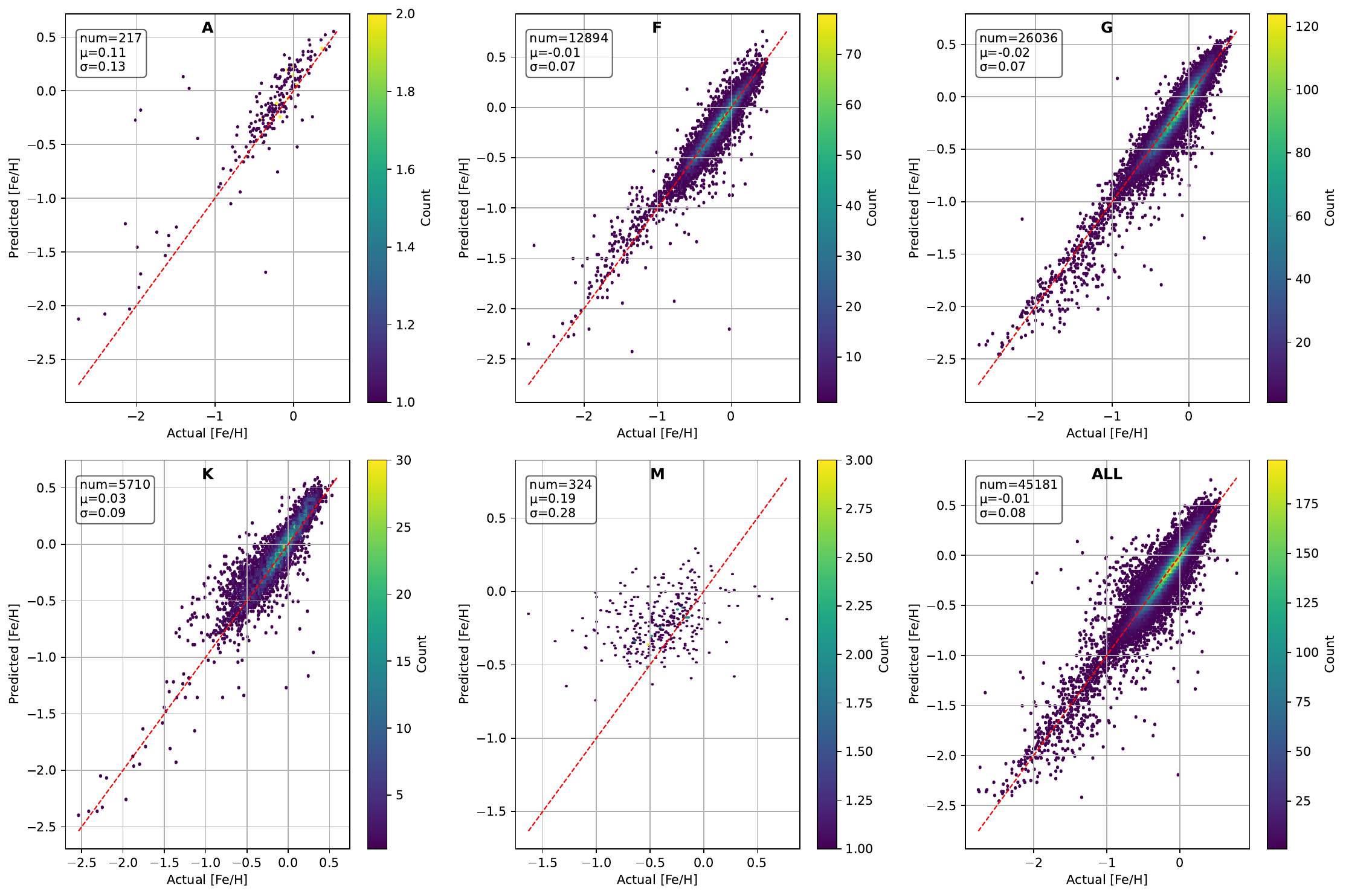}
\caption{[Fe/H] compared with Galah DR4.}
\label{fig:FeH_Galah-DR4}
\end{figure*}

For the \textbf{GALAH DR4} dataset, Figs.~\ref{fig:Teff_Galah-DR4}, \ref{fig:logg_Galah-DR4}, and \ref{fig:FeH_Galah-DR4} present the corresponding external comparisons, with statistical metrics also listed in Table~\ref{tab:Gaia_ESO_galah_dr4_stage_1_model}. Overall, the predicted results for F-, G-, K-, and M-type stars showed good agreement with GALAH DR4 across all three parameters: the point clouds were primarily distributed along the diagonal, and the overall dispersions were small, demonstrating the model’s reliable generalization performance on this dataset.

For $T_{\mathrm{eff}}$, A-type stars displayed relatively larger dispersions. In addition to limitations in model capability, this behavior was also likely related to the intrinsic uncertainties of early-type stellar labels in GALAH DR4 \citep{2025PASA...42...51B}. For $\log g$, the dispersions for A- and M-type stars were slightly higher than those for other spectral types, but remained within an acceptable range, while the remaining spectral types maintained low $\sigma$ values. For [Fe/H], only the M-type subsample showed relatively larger dispersion, whereas the error distributions for other spectral types were more concentrated; this was likely associated with the lower signal-to-noise ratios of late-type stellar spectra.

Taken together, the external comparisons with Gaia-ESO and GALAH DR4 show that the single-stage LLM model maintains consistent performance across different high-resolution spectroscopic surveys. Except for early-type (A-type) or late-type (M-type) stars with limited sample sizes, the parameters $T_{\mathrm{eff}}$, $\log g$, and [Fe/H] exhibit small systematic biases and dispersions for the dominant spectral types (F/G/K), suggesting reliable cross-dataset generalization.

\begin{table*}[htbp]
\centering
\caption{External validation of the first-stage model against Gaia-ESO and GALAH DR4 across stellar spectral subclasses}
\label{tab:Gaia_ESO_galah_dr4_stage_1_model}
\begin{tabular}{lccc|ccc}
\hline
 & \multicolumn{3}{c}{Gaia-ESO} & \multicolumn{3}{c}{GALAH DR4} \\
\cline{2-4} \cline{5-7}
\textbf{Subclass} 
& $T_{\mathrm{eff}}$ ($\mu \pm \sigma$) 
& $\log g$ ($\mu \pm \sigma$) 
& [Fe/H] ($\mu \pm \sigma$)
& $T_{\mathrm{eff}}$ ($\mu \pm \sigma$) 
& $\log g$ ($\mu \pm \sigma$) 
& [Fe/H] ($\mu \pm \sigma$) \\
\hline
A  & $41 \pm 380$    & $0.02 \pm 0.40$  & $-0.06 \pm 0.33$
   & $-280 \pm 186$  & $0.01 \pm 0.23$  & $0.11 \pm 0.13$ \\
F  & $17 \pm 164$    & $0.04 \pm 0.19$  & $-0.02 \pm 0.13$
   & $108 \pm 98$    & $0.03 \pm 0.13$  & $-0.01 \pm 0.07$ \\
G  & $6 \pm 123$     & $-0.06 \pm 0.23$ & $-0.05 \pm 0.15$
   & $28 \pm 86$     & $0.00 \pm 0.14$  & $-0.02 \pm 0.07$ \\
K  & $-63 \pm 128$   & $0.04 \pm 0.17$  & $0.00 \pm 0.18$
   & $-49 \pm 114$   & $0.07 \pm 0.13$  & $0.03 \pm 0.09$ \\
M  & $-102 \pm 147$  & $-0.13 \pm 0.26$ & $0.17 \pm 0.38$
   & $-42 \pm 86$    & $-0.06 \pm 0.24$ & $0.19 \pm 0.28$ \\
\hline
All Classes
   & $-5 \pm 138$    & $-0.02 \pm 0.22$ & $-0.03 \pm 0.16$
   & $42 \pm 105$    & $0.02 \pm 0.14$  & $-0.01 \pm 0.08$ \\
\hline
\end{tabular}
\end{table*}

\subsubsection{Comparison between the Second-Stage Model and External Datasets}

\begin{figure*}[t]
\centering
\includegraphics[
  width=\textwidth,
  height=0.98\textheight,
  keepaspectratio
]{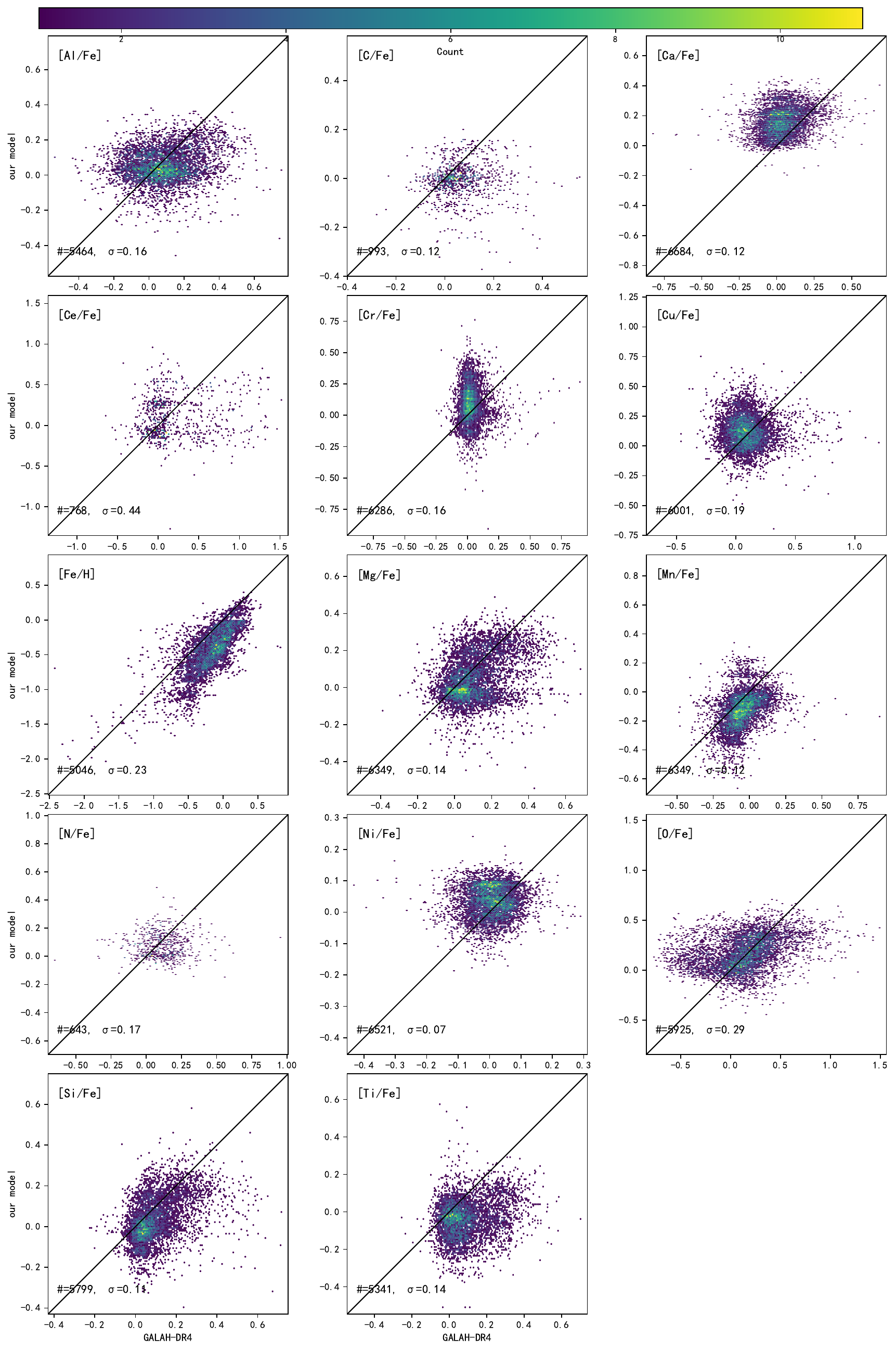}
\caption{Abundance compared with Gaia ESO.}
\label{fig:Abundance_Gaia_ESO}
\end{figure*}

\begin{figure*}[t]
\centering
\includegraphics[
  width=\textwidth,
  height=0.98\textheight,
  keepaspectratio
]{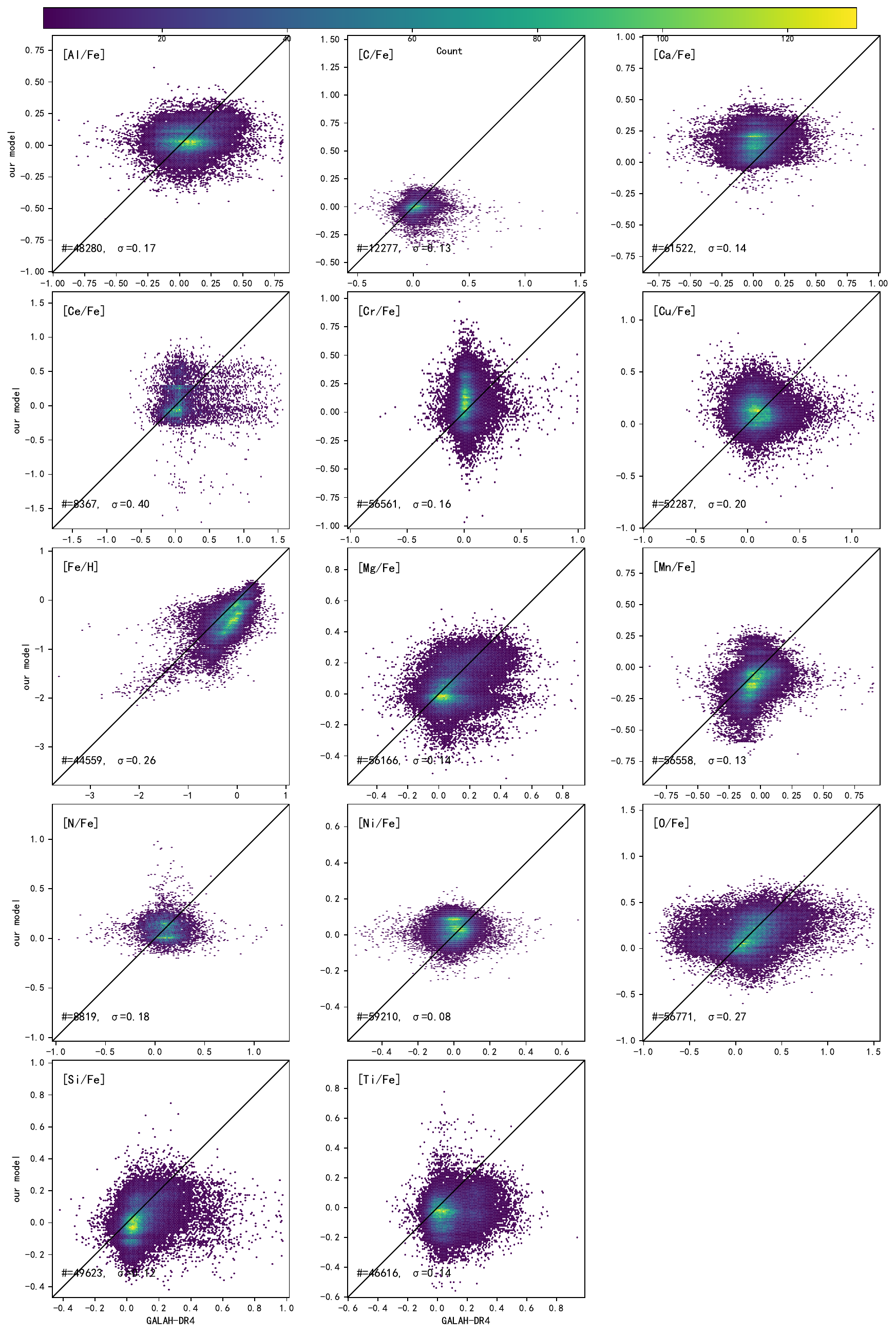}
\caption{Abundance compared with GALAH DR4.}
\label{fig:Abundance_GALAH_DR4}
\end{figure*}

In the previous subsection, we have verified the external consistency of the single-stage LLM model. Here, we further extend the analysis to the two-stage model by systematically comparing its element-by-element abundance predictions with those from the external spectroscopic surveys Gaia-ESO and GALAH DR4. As shown in Table~\ref{tab:Gaia_ESO_galah_dr4_stage_2_model}, we assess the reliability of the two-stage model at the chemical abundance level by examining the degree of consistency for multiple elements across different datasets. Figures~\ref{fig:Abundance_Gaia_ESO} and~\ref{fig:Abundance_GALAH_DR4} present one-to-one correlation plots for a subset of elements, comparing the model-inferred results with stellar parameters derived from high-resolution spectra in \textit{Gaia-ESO} and \textit{GALAH DR4}, respectively.

Overall, the conclusions drawn from the two external datasets are broadly consistent. Most elements exhibit a trend in which the scatter clouds approach the 1:1 dashed line in both comparisons. However, substantial differences are observed among individual elements: one group shows small scatter or low dispersion (indicating good external transferability), while another group displays much larger scatter and more complex distribution patterns, reflecting weaker external consistency.

In the \textbf{Gaia-ESO} comparison (Table~\ref{tab:Gaia_ESO_galah_dr4_stage_2_model}), the model demonstrates the most stable external transferability for a set of elements, including
\texttt{MG\_FE}, \texttt{SI\_FE}, \texttt{CA\_FE}, \texttt{TI\_FE}, \texttt{MN\_FE}, \texttt{NI\_FE}, \texttt{C\_FE}, and \texttt{N\_FE}.
In the corresponding scatter-density plots (Fig.~\ref{fig:Abundance_Gaia_ESO}), the point clouds for these elements form relatively narrow bands along the 1:1 relation, with generally small dispersions $\sigma$ (e.g., \texttt{NI\_FE} exhibits the smallest $\sigma$, while the other elemental abundances also maintain low scatter). This indicates a high degree of consistency and robustness between the model predictions and the external survey parameters for these elements. Meanwhile, \texttt{AL\_FE}, \texttt{K\_FE}, \texttt{CR\_FE}, and \texttt{CU\_FE} show intermediate levels of dispersion: although their scatter distributions still broadly follow the 1:1 relation, the bandwidth is wider, suggesting that they may be adequate for studies of global chemical trends but should be used with greater caution in fine-structure analyses.

In contrast, the elements with relatively weaker external consistency in the \textit{Gaia-ESO} comparison are mainly
\texttt{O\_FE}, \texttt{NA\_FE}, \texttt{V\_FE}, \texttt{CO\_FE}, and \texttt{CE\_FE}.
Among them, \texttt{O\_FE}, \texttt{NA\_FE}, and \texttt{V\_FE} exhibit significantly increased dispersion (with $\sigma$ values markedly higher than those of the stable element group), accompanied by more diffuse scatter distributions and more pronounced deviations from the 1:1 trend. \texttt{CO\_FE} not only shows relatively large dispersion, but also presents a clear negative zero-point bias (with a negative and relatively large $\mu$), and is more prone to non-linear structures. \texttt{CE\_FE} displays the largest dispersion and the weakest external consistency.

\begin{table}[t]
\centering
\caption{External validation of elemental abundance predictions for the second-stage model against Gaia-ESO and GALAH DR4}
\label{tab:Gaia_ESO_galah_dr4_stage_2_model}
\footnotesize
\begin{tabular}{lrrr!{\vrule width 0.8pt}rrr}
\toprule
& \multicolumn{3}{c!{\vrule width 0.8pt}}{\textbf{Gaia-ESO}} & \multicolumn{3}{c}{\textbf{Galah DR4}} \\
\cmidrule(r{0.5em}){2-4} \cmidrule(l{0.5em}){5-7}
Elem. & $N$ & $\mu$ & $\sigma$ & $N$ & $\mu$ & $\sigma$ \\
\midrule
O\_FE  &  5925 &  0.03  & 0.27          & 56771 & $-$0.01 & 0.25 \\
NA\_FE &  6620 &  0.03  & 0.25          & 60817 &  0.01  & 0.26 \\
MG\_FE &  6349 & $-$0.05 & \textbf{0.13} & 56166 & $-$0.04 & \textbf{0.13} \\
AL\_FE &  5464 & $-$0.03 & \textbf{0.16} & 48280 & $-$0.03 & \textbf{0.16} \\
SI\_FE &  5582 & $-$0.04 & \textbf{0.10} & 49623 & $-$0.06 & \textbf{0.10} \\
K\_FE  &  6510 &  0.15  & \textbf{0.19} & 60117 &  0.16  & \textbf{0.20} \\
CA\_FE &  6684 &  0.12  & \textbf{0.12} & 61522 &  0.13  & \textbf{0.13} \\
TI\_FE &  5341 & $-$0.13 & \textbf{0.13} & 46616 & $-$0.12 & \textbf{0.13} \\
V\_FE  &  5481 &  0.09  & 0.28          & 47977 &  0.09  & 0.27 \\
CR\_FE &  6016 &  0.07  & \textbf{0.15} & 56561 &  0.07  & \textbf{0.15} \\
MN\_FE &  6007 & $-$0.08 & \textbf{0.11} & 56558 & $-$0.07 & \textbf{0.12} \\
CO\_FE &  5271 & $-$0.27 & 0.24          & 46105 & $-$0.24 & 0.21 \\
NI\_FE &  6430 &  0.03  & \textbf{0.07} & 59210 &  0.02  & \textbf{0.08} \\
CU\_FE &  5721 &  0.03  & \textbf{0.18} & 52287 &  0.02  & \textbf{0.18} \\
C\_FE  &  2055 & $-$0.06 & \textbf{0.09} & 12277 & $-$0.04 & \textbf{0.10} \\
N\_FE  &  1144 & $-$0.03 & \textbf{0.15} &  8819 & $-$0.02 & \textbf{0.17} \\
CE\_FE &  1139 & $-$0.09 & 0.32          &  8367 & $-$0.03 & 0.34 \\
\bottomrule
\end{tabular}
\end{table}

In the \textbf{GALAH DR4} comparison (Table~\ref{tab:Gaia_ESO_galah_dr4_stage_2_model}), the above grouping is reaffirmed. The elements with the best external consistency remain
\texttt{MG\_FE}, \texttt{SI\_FE}, \texttt{CA\_FE}, \texttt{TI\_FE}, \texttt{MN\_FE}, \texttt{NI\_FE}, \texttt{C\_FE}, and \texttt{N\_FE},
all of which maintain low $\sigma$ values under both external reference scales, with scatter patterns most closely aligned with the 1:1 relation, indicating strong external consistency. \texttt{AL\_FE}, \texttt{K\_FE}, \texttt{CR\_FE}, and \texttt{CU\_FE} continue to form a group characterized by moderate dispersion but relatively consistent trends. The elements with weaker external consistency are again concentrated in
\texttt{O\_FE}, \texttt{NA\_FE}, \texttt{V\_FE}, \texttt{CO\_FE}, and \texttt{CE\_FE}:
among these, \texttt{CE\_FE} still exhibits the largest dispersion in GALAH DR4, while \texttt{CO\_FE} continues to show a pronounced negative $\mu$.

By combining the results from both external reference scales, we classify the external performance of the elements into three categories:

(i) \emph{Well-determined and stable}: \texttt{MG\_FE}, \texttt{SI\_FE}, \texttt{CA\_FE}, \texttt{TI\_FE}, \texttt{MN\_FE}, \texttt{NI\_FE}, \texttt{C\_FE}, and \texttt{N\_FE};

(ii) \emph{Moderately stable (usable but with slightly reduced precision)}: \texttt{AL\_FE}, \texttt{K\_FE}, \texttt{CR\_FE}, and \texttt{CU\_FE};

(iii) \emph{Relatively weak external consistency}: \texttt{O\_FE}, \texttt{NA\_FE}, \texttt{V\_FE}, \texttt{CO\_FE}, and \texttt{CE\_FE}.

From both physical and data-processing perspectives, the latter group of elements can be often more limited by factors such as an insufficient number of effective diagnostic lines, weak or blended features, and differences among surveys in non-LTE/3D treatments, line lists, and prior assumptions. As a result, they are more likely to exhibit larger scatter or systematic offsets in external comparisons. Accordingly, for studies of chemical evolution trends, population classification, or chemical tagging, we recommend prioritizing robust tracers such as \texttt{MG\_FE}, \texttt{SI\_FE}, \texttt{CA\_FE}, \texttt{TI\_FE}, \texttt{MN\_FE}, \texttt{CR\_FE}, \texttt{NI\_FE}, \texttt{C\_FE}, and \texttt{N\_FE}. For \texttt{O\_FE}, \texttt{NA\_FE}, \texttt{V\_FE}, \texttt{CO\_FE}, and \texttt{CE\_FE}, stricter quality cuts are advised, along with explicit correction or modeling of systematic effects, or the inclusion of larger systematic uncertainties in scientific interpretations.

\subsection{Limitations and Future Work}

Despite its advantages, the LLM still has limitations when applied to spectroscopic tasks, and we outline possible directions for improvement. Specifically, for the prediction of the three fundamental atmospheric parameters, the remaining shortcomings are mainly manifested in two aspects: (i) there is still room for improvement in the prediction of $T_{\mathrm{eff}}$ and $\log g$ for early-type A stars, where the generalization capability needs to be further strengthened; and (ii) the prediction of $T_{\mathrm{eff}}$ for late-type M stars also requires additional improvement. In terms of underlying causes, these limitations primarily are likely related to insufficient precision in capturing weak and blended spectral features, as well as inadequate feature learning for a small number of underrepresented spectral types.

To address the above issues, future work can mitigate these limitations through a series of targeted strategies. For example, additional samples of specific spectral types such as A and M stars can be incorporated for dedicated training to enhance the model’s learning of underrepresented classes; the model architecture can be optimized to strengthen the extraction mechanisms for weak and blended spectral lines; and model ensemble strategies can be introduced to combine the feature extraction strengths of traditional models with the global information integration capability of LLMs, to leverage complementary strengths. Through these improvements, the application potential of LLMs in spectroscopic tasks could be further improved, facilitating broader and more effective applications in astronomical spectral data analysis.

In element-by-element abundance prediction tasks, although the LLM demonstrates clear advantages for several key elements, its overall performance has not yet comprehensively surpassed that of traditional models across all elemental dimensions. Compared with fundamental atmospheric parameters, element-wise abundance regression is more sensitive to the physical origins of spectral lines, the precision of local feature representation, and the degeneracy induced by line blending. Consequently, the model is required not only to understand the global contextual structure of spectra, but also to achieve high-precision characterization of individual or a small number of spectral features within specific wavelength regions. The performance variation of a single modeling paradigm across different elements therefore may reflects an inherent trade-off between the global modeling strengths of LLMs and the fine-grained fitting of local spectral lines.

Building on this complementary nature, future work may further explore hybrid modeling strategies tailored for element-wise abundance tasks. For instance, convolutional or physics-informed prior modules can be introduced for elements dominated by strong local spectral lines, while the contextual modeling capability of LLMs can be enhanced for elements that rely on the synergy of multiple lines and cross-band information. In this way, an overall improvement in element-by-element abundance inversion accuracy may be achieved within a unified framework.

\section{Conclusion} \label{sec:conclusion}

In this study, we propose a large language model (LLM) framework for stellar spectral parameter inversion, in which stellar spectra are modeled as language-like sequences and a \emph{two-stage training} strategy is adopted. Specifically, the model first learns general spectral representations from large-scale low-resolution spectra from LAMOST, and is then fine-tuned using high-quality labels from APOGEE to enable joint prediction of multiple physical quantities. Experimental results show that the proposed approach can accurately and jointly infer three fundamental atmospheric parameters ($T_{\rm eff}$, $\log g$, and $[\mathrm{Fe/H}]$), and can also jointly predict the abundances of nearly twenty chemical elements. Compared with traditional machine learning methods and convolutional neural network baselines, the LLM exhibits smaller error dispersion and more consistent statistical behavior across most spectral types. Notably, it maintains improved robustness even in low signal-to-noise regimes or subsamples with imbalanced distributions, highlighting its potential for large-scale survey data processing.

Importantly, based on comparative experiments with data scales of $10^5$, $5\times 10^5$, and $1.5\times 10^6$, we observe a \emph{scaling} trend in model performance. As the training sample size and effective model capacity increase, the systematic biases ($\mu$) of the three fundamental parameter predictions consistently converge toward zero, while the dispersions ($\sigma$) are further reduced, yielding more uniform error distributions across major spectral types. This behavior is consistent with scaling laws observed in natural language tasks, indicating that the paradigm of modeling “spectra as language” possesses favorable scalability. Furthermore, cross-validation against external high-resolution surveys such as Gaia-ESO and GALAH DR4 shows that the model preserves good physical plausibility across datasets. For F/G/K-type stars, both the systematic biases and dispersions of $T_{\rm eff}$, $\log g$, and $[\mathrm{Fe/H}]$ remain relatively small, while at the elemental abundance level, several key elements exhibit stable agreement across different external reference scales, further lending support to the reliability of the model predictions.

Nevertheless, early-type A stars and late-type M stars remain challenging for both atmospheric parameter regression and element-by-element abundance inference. In addition, the external consistency of some elements is relatively weaker likely due to factors such as intrinsically weak or blended spectral lines, a limited number of effective diagnostic features, and differences in line lists and prior treatments among surveys. Future work may further improve performance from the perspectives of data, model architecture, and training strategy. Possible directions include targeted data augmentation and rebalanced training for A/M-type samples, the incorporation of physics-informed feature enhancement mechanisms tailored to weak-line and molecular-band regions, and the exploration of hybrid or integrated frameworks combining LLMs with CNNs to jointly leverage global contextual modeling and fine-grained local spectral feature extraction within a unified system. In summary, this work shows the feasibility and potential of large language models in quantitative spectroscopy, and provides a scalable new technical pathway for efficient and precise stellar parameter determination in forthcoming larger-scale and higher-resolution spectroscopic surveys.

\section{Funding}
This work is supported by the National Natural Science Foundation of China (12273078, 12411530071, 12273075), National Key Research and Development Program（Grant No. 2025YFF0510602）and National Astronomical Observatories of the Chinese Academy of Sciences (No.E4ZR0516). 

\bibliography{reference}{}

\bibliographystyle{aasjournal}

\end{document}